\documentclass[twocolumn,epsfig,pre]{revtex4}

\usepackage{graphics}
\usepackage{graphicx}
\usepackage{epstopdf}
\usepackage{amssymb}
\usepackage{amsmath}
\usepackage{hyperref}
\usepackage{latexsym}

\begin{document}

\renewcommand{\thefootnote}{\fnsymbol{footnote}}
\renewcommand{\theequation}{\arabic{section}.\arabic{equation}}

\title{Particle dynamics in fluids with random interactions}

\author{Lenin S. Shagolsem}
\email{slenin2001@gmail.com}
\author{Yitzhak Rabin}
\email{yitzhak.rabin@biu.ac.il}
\affiliation{Department of Physics, and Institute of Nanotechnology and Advanced Materials,
Bar-Ilan University, Ramat Gan 52900, Israel}

\date{\today}

\begin{abstract}

\noindent {We study the dynamics of particles in a multi-component 2d Lennard-Jones (LJ) fluid in the limiting case where {\it all the particles are different} (APD). The equilibrium properties of this APD system were studied in our earlier work [J.~Chem.~Phys.~142, 051104]. We use molecular dynamics simulations to investigate the statistical properties of particle trajectories in a temperature range covering both normal and supercooled fluid states. We calculate the mean-square displacement as well as angle, displacement and waiting time distributions, and compare the results with those for one-component  LJ fluid. As temperature is lowered, the dynamics of the APD system becomes increasingly complex, as the intrinsic difference between the particles is amplified by neighborhood identity ordering and by the heterogeneous structure of the supercooled state.}
\end{abstract}

\maketitle


\section{Introduction}
\label{sec: introduction}

The tracking of individual particles and the study of their motions provides a simple means to probe the dynamics of complex fluids, either experimentally or theoretically (via computer simulations). The statistical analysis of particle trajectories can be used to uncover the physical origins of the broad distributions of relaxation times and the heterogeneous dynamics (e.g., in particle displacement distributions) observed in glass formers and colloidal suspensions close to glass and jamming transitions, respectively \cite{SSastry1998_Nature,ERWeeks2000_Science,ERWeeks2002_PRL,Nagel_Nature_2005,PinakiC2007_PRL,
HTanaka2010_NatureMaterials,JHelfferich2015_PRE1,Igal}, and 
in simple one-component fluids near the freezing transition in two dimensions \cite{AZPatashinski2012_JPCL}.  
For example, computer simulation studies of glass-forming liquids suggest that (a) the slow dynamics and medium-range crystalline order are correlated \cite{TKawasaki2007_PRL}, (b)  heterogeneous dynamics is a consequence of critical-like fluctuations of static structural order \cite{HTanaka2010_NatureMaterials}, and (c) structural and dynamical heterogeneity are correlated ~\cite{Matharoo2006_PRE}. Furthermore, particle tracking experiments on colloids~\cite{ERWeeks2000_Science,ERWeeks2002_PRL} and theory/simulation studies of glass-formers and colloids (see e.g.~ref.~\cite{PinakiC2007_PRL} and references therein) reveal that such complicated dynamical features are associated (at the single particle level)  with intermittent particle motion- confinement of particle in cage formed by the neighboring particles and its subsequent release (trapped and jump-like motion);  sting-like cooperative motion of particles is also observed~\cite{ERWeeks2002_PRL,CDonati1998_PRL}.\ 

Complex dynamics is also very common in crowded and heterogeneous bio-environments (e.g., living cells), as revealed by the single particle tracking experiments \cite{eli2012_PhysToday,MRHorton2010_SoftMatter,MJSaxton1997_Biophys,IIzeddin2014_eLife,SRMcGuffee2010_PLoS_ComputBiol}. In living cells, many of the constituents (e.g., proteins) differ in shape, size and in their interactions with each other and with the surrounding medium; furthermore,  they can be active (i.e., generate forces by converting chemical energy into mechanical work, upon ATP hydrolysis). There are several models of anomalous diffusion in biological systems, e.g., transient binding/unbinding, membrane compartmentalization, and membrane heterogeneity~\cite{IMSokolov2012_SoftMatter} (for a review on models of anomalous diffusion in crowded environments, see ref.\cite{RMetzler2000_PhysicsReports}). 

In the present study we consider a system of intermediate level of complexity between the former (glass-formers and colloids) and the latter systems (proteins in cells), and focus on the dynamics of particles with different pair interaction strengths but with no size/shape polydispersity, that interact with each other and experience random forces of thermal origin only. We will carry out a comprehensive statistical analysis of the dynamics of different types of particles in a multi-component 2d Lennard-Jones (LJ) fluid in the limiting case where {\it all the particles are different} (APD). Note that these particles differ only in the strength of their attractive interactions with other particles and since these attractions  become increasingly important as temperature is lowered, in this work we will focus on the liquid to solid transition region, i.e. on the range of temperatures at which the differences between the particles are amplified and at the same time the system is not yet kinetically frozen. We will show that the dynamics of different particles in an APD system depends on their interaction parameters. In the fluid phase above the freezing transition temperature the motion of all the particles is diffusive and the only difference is in the magnitude of their diffusion coefficients; strongly interacting particles diffuse slower than weakly interacting ones. Qualitative differences between the motions of different types of particles appear in the supercooled range below the transition temperature where the system becomes strongly heterogeneous (crystalline domains with mobile defects coexist with disordered low-density fluid regions) and where the dynamic consequences of microphase separation between strongly and weakly interacting particles are revealed.

Our paper is organized as follows: In section~\ref{sec: model-simulation-details}, we present the model and 
the simulation details and briefly describe the main results of our previous work on equilibrium properties of APD systems.
The analysis of mean square displacement, angle,
displacement, and waiting time distributions is presented in
sections~\ref{sec: msd-analysis}~-~\ref{sec: waiting-time-analysis}. In
section~\ref{sec: conclusion} we summarize and discuss the results of this work.

\section{Model and simulation details}
\label{sec: model-simulation-details}

We perform MD simulations (in NVT-ensemble) in two dimensions with $N=2500$ particles in a simulation box of size $L_x=L_y=60\sigma$, where $\sigma$ is the particle diameter. At this number density $\rho^\ast=0.6944~(=N/L_xL_y)$ the system behaves as a liquid at sufficiently high temperatures. Periodic boundary conditions are applied along both x and y directions. All the particles have same size $\sigma$ and mass $m$ which are set to unity. Particles in the system interact via Lennard-Jones (LJ) potential
\begin{equation}
U_{ij}(r) = 4\epsilon_{ij}\left[(\sigma/r)^{12}-(\sigma/r)^{6}\right]~,
\label{eqn: LJ-potential}
\end{equation}
where $\epsilon_{ij}$ and $r$ are the interaction strength and separation between a pair of particles $i$ and $j$ respectively. The potential is truncated and shifted to zero at $r=2.5\sigma$.\

\noindent The equation of motion of  particle $i$ is given by the Langevin equation
\begin{equation}
 m \frac{d^2 \bf r_i}{dt^2} + \zeta \frac{d \bf r_i}{dt} = -\frac{\partial U}{\partial \bf r_i}
 + {\bf f_i}~,
 \label{eqn: langevin}
\end{equation}
with $\bf r_i$ the position of particle $i$, $\zeta$ the friction coefficient (assumed to be the same for all particles), $U$ is the sum of all the pair potentials
$U_{ij}$ ($i\ne j$) for a given spatial configuration of the system,
and $\bf f_i$ a random external force with zero mean and second moment
proportional to the poduct of temperature $T$ and $\zeta$. All the physical quantities are expressed in
LJ reduced units, with LJ time $\tau_{_{\rm LJ}}=1$ (the MD simulation time step used in the integration of the equations of motion
is 0.005) \cite{allen}.
The friction coefficient can written as $\zeta=1/\tau_d$, with $\tau_d$ the characteristic viscous damping time which
we fixed at $50$ and which determines the transition from inertial to overdamped motion (due to collisions with molecules of the implicit "solvent") in the dilute system limit. At the rather high density in the present simulation, the damping due to collisions between the particles dominates and the transition to the overdamped regime takes place on much faster time scales (of the order of 0.1). The simulations were carried out using LAMMPS code\cite{lammps}.

\begin{figure}[ht]
\includegraphics*[width=0.4\textwidth]{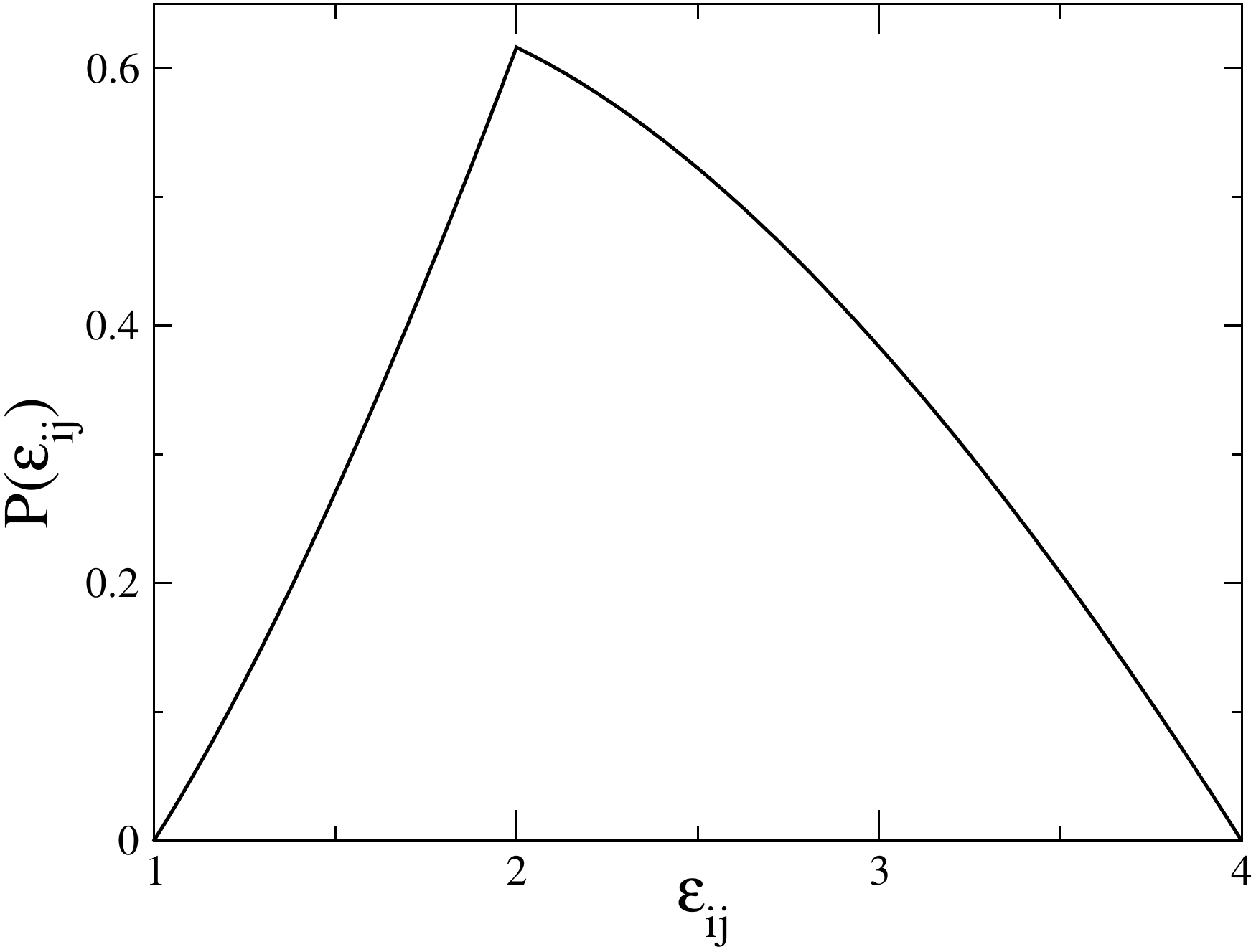}
\begin{center}
\caption{Distribution of pair interaction parameters $P(\epsilon_{ij})$ in the range $1-4$ for
GM system.}
\label{fig: epsij-distribution}
\end{center}
\end{figure}

In order to characterize the interactions between the different particles, we have to introduce a distribution of pair interaction parameters $\epsilon_{ij}$ that enter the LJ potential in Eq. \ref {eqn: LJ-potential}. In this work we follow the geometric mean (GM) prescription, introduced in ref. \cite {Shagolsem2015_JCP} according to which a particle $i$ is assigned an interaction strength $\epsilon_i$ taken randomly from a uniform distribution  in the range 1-4  (note that the identity of each particle is  defined by its interaction strength and since the latter is a random number, all particles are different from each other). The pair interaction parameter between particles $i$ and $j$ is defined as the geometric mean of their interaction strengths , i.e., $\epsilon_{ij}=\sqrt{\epsilon_i\epsilon_j}$. This leads to a peaked distribution of the pair interaction parameters $P(\epsilon_{ij})$ with most probable value $\epsilon_{ij}^{max}=2.0$ and mean $\langle\epsilon_{ij}\rangle=2.42$ (fig.~\ref{fig: epsij-distribution}).
As a reference system, we simulate a one-component (1C) LJ system with the same
density as the GM system and with pair interaction parameter $\epsilon_{ij}=2.5$, the mid-point of the interval $1-4$.\\

The static properties of the GM system (as well as other probability distribution functions of pair interaction parameters) have been studied in detail in ref. \cite {Shagolsem2015_JCP}. The main finding of the above work was that as the system is cooled from the high temperature regime in which kinetic energy dominates over the attractive potential (and thus, the difference between the particles becomes unimportant), the GM system undergoes a smooth transition into a self-organized fluid state characterized by short-range {\it neighborhood identity ordering} (NIO). In this state the identities of neighboring particles become strongly correlated in the sense that particles with high values of $\epsilon_i$ tend to cluster together (and similarly for particles with low values of $\epsilon_i$), resulting in microphase separation between particles with high and low interaction strengths.  NIO becomes increasingly more pronounced with decreasing temperature, a trend that continues until the system becomes kinetically frozen.

It was found that the liquid-solid transition takes place within a narrow temperature interval in which the mean interaction energy changes rapidly upon cooling from the liquid to the ``supercooled fluid'' (using the terminology of ref. \cite{HTanaka2010_NatureMaterials}) state,  in which a hexagonally-ordered solid with mobile defects coexists with a rarefied fluid of particles moving inside voids in the solid. The transition temperature  can be defined as the temperature at which the radial distribution function decays over half the system size or by using other criteria. All methods yield $T^\ast \approx 1.0$ for both the GM and the 1C systems (at the same density as in the present work). The precise nature of this transition is still controversial even for 1C systems and may be affected by finite size effects \cite {mitus, patashinsky}. Both the static and the dynamic properties of the GM and 1C systems are expected to be quite similar for $T\gg T^ \ast$ and differences between the two systems become significant as the temperature is decreased, approaching the upper limit ($\epsilon_{ij}=4$) of pair interaction parameter values in the GM system, and neighborhood identity ordering becomes increasingly important (there is no NIO  in the 1C system). Since the differences between the two systems are amplified as the transition temperature ($T^\ast=1$) is approached, in the present study we focus on the temperature range around the transition temperature and compare the GM and 1C systems at the same value of the reduced temperature $\delta = (T-T^\ast)/T^\ast$.  Note that the supercooled fluid regime extends from the equilibrium transition temperature  ($\delta =0$) down to the
temperature ( $\delta\approx ~-0.5$ for both the 1C and the GM systems) at which all the particles have condensed into the ordered solid phase and all voids and lattice defects became immobile on our simulation time scales.

The longest time intervals in our simulations were $2\times 10^5\tau_{LJ}$ which corresponds to 40 million MD time steps. All the results reported in this work were obtained after the systems were equilibrated in the sense that ensemble-averaged properties (e.g., mean potential energy per particle) were time-independent and no further aging was observed. In order to get better statistics, all the distribution functions were calculated by collecting data along each trajectory and adding up data from the trajectories of all the particles in the particular ensemble studied. Note that averages obtained in this way involve both time and ensemble averaging and that the question of ergodicity (the difference between time and ensemble averages) did not arise. Nevertheless, we can not  exclude the possibility of weak ergodicity breaking effects associated with probing individual particle trajectories over finite time intervals \cite {barkai, manzo}.  The study of such effects will be the subject of future work.



\section{Mean square displacement}
\label{sec: msd-analysis}

The simplest measure of the statistical properties of particle trajectories is the mean square displacement (MSD) during time $t$. We begin by  calculating the MSD averaged over all the particles
\begin{equation}
 \left\langle \Delta r^2 (t) \right\rangle = \frac{1}{N}\sum_{i=1}^N \left(r_i(t)-r_i(0)\right)^2~,
 \label{eqn: MSD}
\end{equation}
 with $i$ the particle index.

\begin{figure}[ht]
\includegraphics*[width=0.45\textwidth]{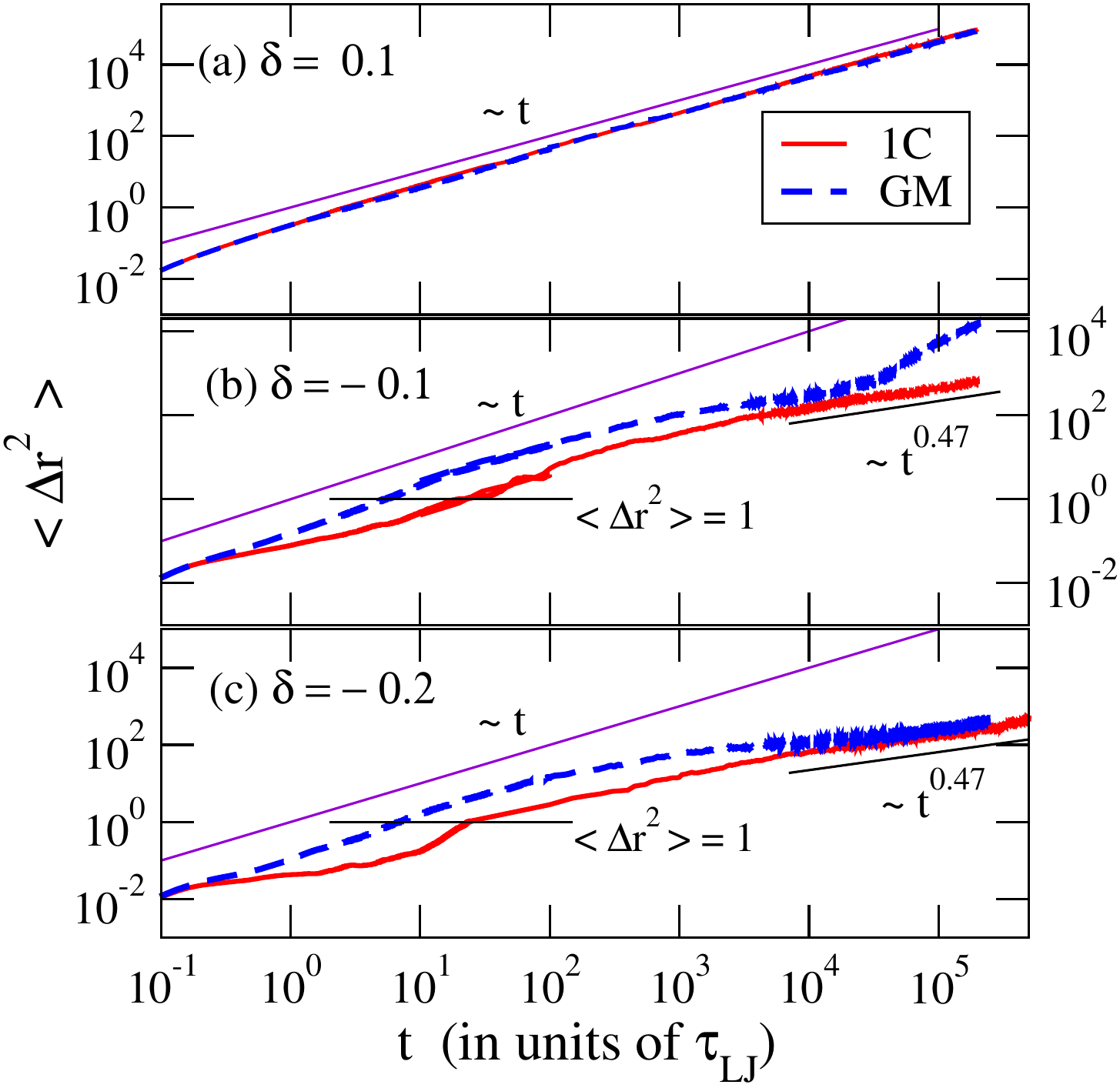}
\begin{center}
\caption{ MSD for 1C and GM (the latter is averaged over all particle types) systems shown at three different values of $\delta$ indicated in the figure. Horizontal line shows $\left\langle \Delta r^2 \right\rangle = 1$ (particle size).}
\label{fig: msd-compare}
\end{center}
\end{figure}
In figure~\ref{fig: msd-compare} we compare the MSD for 1C and GM systems (since all particles are different in the GM system, the average over all the particles means also that we average over all particle types), at temperatures both above and below $T^\ast$.  Inspection of fig. \ref{fig: msd-compare}(a) shows that in agreement with expectation of simple diffusion in the normal fluid state, at temperatures above the transition $\delta = 0.1$, the MSD grows linearly with time for both systems. The temperature dependence of the diffusion coefficient in both systems for $T>T^\ast$
follows Arrhenius-like behavior and is shown in the SI. Note that the two MSD curves lie on top of each other above the transition temperature indicating that the corresponding diffusion coefficients are nearly identical. 

The difference between the MSDs in the 1C and the GM systems is resolved (the MSD for the 1C system is always lower than that for the GM system) at temperatures below the transition, $\delta = -0.1$ and $\delta = -0.2$, where attractive interactions dominate over the kinetic energy and the effects of a broad distribution of interaction parameters survive the averaging over the particle types. Interestingly, as temperature is lowered through the supercooled range, the MSD curve in the 1C system develops a plateau at short times, followed by a transition to a subdiffusive regime ($MSD\propto t^{1/2}$) at longer times. Upon some reflection we conclude that the plateau describes the vibration of particles inside solid clusters with amplitude that corresponds to about $10\%$ of the interparticle separation (i.e., below the Lindemann criterion for melting), followed by escape from the ``cage'' at longer times, when the MSD reaches the particle diameter (the solid line in figs. \ref{fig: msd-compare} (b) and (c)) and finally cross-over to subdiffusion, presumably due to the presence of mobile vacancies and defects. No short-time plateau is observed for the GM system, where the above-mentioned effects are smeared by the heterogeneous kinetics of different particle types.  At very long times $t>10^4$,  superdiffusive  behavior is observed for the GM system at intermediate temperature, $\delta = -0.1$ (see fig. \ref{fig: msd-compare}(b)), possibly signifying a transition to yet another regime that lies outside the time scales reached in our simulation. The difference between the MSDs of the two systems disappears in the long time limit at yet lower temperature, $\delta = -0.2$ (see fig. \ref{fig: msd-compare}(c)).

\begin{figure}[ht]
\includegraphics*[width=0.5\textwidth]{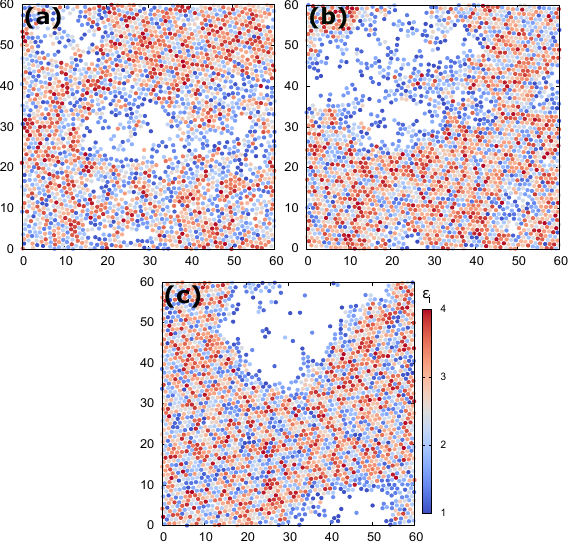}
\begin{center}
\caption{Typical spatial ordering of particles in GM system where the particles are colored according to their $\epsilon_i$ values,  for (a) $\delta=0$, (b) $\delta= -0.1$, and (c) $\delta= -0.2$. Color bar indicates the magnitude of $\epsilon_i$ on a continuous scale from blue (1) to red (4).}
\label{fig: config-gm-epsi}
\end{center}
\end{figure}

\begin{figure}[ht]
\includegraphics*[width=0.45\textwidth]{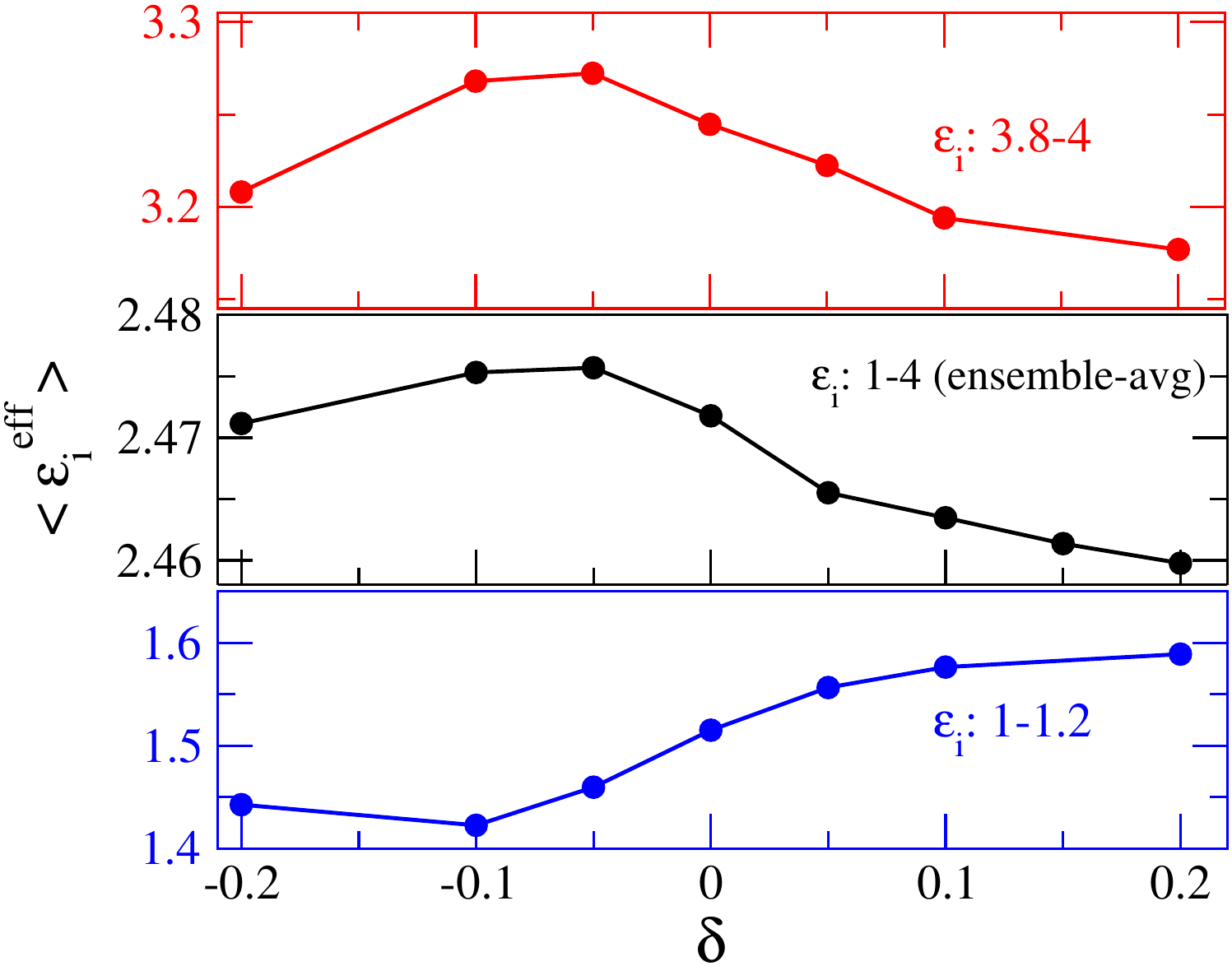}
\begin{center}
\caption{Variation of $\left\langle \epsilon_i^{\rm eff} \right\rangle$ with $\delta$, shown for two different groups of particles with values of $\epsilon_i$ in the ranges $1-1.2$ (upper panel) and $3.8-4$ (lower panel), and averaged over all particles in GM system (middle panel).}
\label{fig: eeff-mean-gm}
\end{center}
\end{figure}

We now proceed to investigate the effects of particle identity (value of $\epsilon_i$) on its dynamics in different temperature regimes.  Typical configurations of the GM system at ($\delta=0$) and below ($\delta=-0.1$ and $-0.2$) the liquid-solid transition are shown in figures \ref{fig: config-gm-epsi} (a) - \ref{fig: config-gm-epsi}(c) where the color code indicates high (red) and low (blue) values of  $\epsilon_i$. Inspection of these figures shows that particles with larger
values of $\epsilon_i$ tend to lie inside solid clusters, while particles inside voids and at the boundaries and the defects of the solid are predominantly of  smaller $\epsilon_i$ type. 

Neighborhood identity ordering is manifested by the magnitude of the effective interaction parameter of particle $i$, defined as
\begin{equation}
\epsilon_i^{\rm eff}=\frac {1}{n_b}\sum_{j=1}^{n_b}\epsilon_{ij}~,
 \label{eqn: <epsilon>}
\end{equation}

\noindent with $n_b$ the number of nearest neighbor particles\cite{Shagolsem2015_JCP}. In fig. \ref{fig: eeff-mean-gm} we plot the ensemble average of this parameter, $\left\langle \epsilon_i^{\rm eff} \right\rangle$ as a function of temperature,  where the averaging is either over all the particles in the GM system (middle panel), or over a given subset of all particles (upper and lower panels). The increasing splitting between the  $\left\langle \epsilon_i^{\rm eff} \right\rangle$ values of high and low $\epsilon_i$ particles  shown in the upper and the lower panels, respectively, is the signature of NIO: the tendency of particles with similar values of $\epsilon_i$ to cluster together leads to increase (decrease) of $\left\langle \epsilon_i^{\rm eff} \right\rangle$ for high (low) $\epsilon_i$ particles (at high temperatures NIO vanishes and  $\left\langle \epsilon_i^{\rm eff} \right\rangle \rightarrow \sqrt{2.5\epsilon_i}$, in agreement with our simulation results). Surprisingly, fig. \ref{fig: eeff-mean-gm} (middle panel) shows that NIO is a non-monotonic function of temperature below the liquid-solid transition, and {\it decreases} as temperature is lowered beyond
$\delta\approx -0.1$. Inspection of figs.  \ref{fig: config-gm-epsi}(b) and (c) suggests this non-monotonic behavior is the consequence of condensation of (low $\epsilon_i$) particles of the fluid on the solid clusters (composed of higher $\epsilon_i$ particles) as $\delta$ decreases from $-0.1$ to $-0.2$. Since the value of $\left\langle \epsilon_i^{\rm eff} \right\rangle$ depends on the local environment of the particle, this leads to the observed increase of $\left\langle \epsilon_i^{\rm eff} \right\rangle$ for the low $\epsilon_i$ particles (lower panel in fig. \ref{fig: eeff-mean-gm}) and to the decrease of  $\left\langle \epsilon_i^{\rm eff} \right\rangle$ for the \
high $\epsilon_i$ particles (upper panel in fig. \ref{fig: eeff-mean-gm}). The balance between the two effects is delicate but inspection of the middle panel in fig. \ref{fig: eeff-mean-gm} shows that the latter effect dominates over the former one in the above temperature range.
\begin{figure}[ht]
\includegraphics*[width=0.45\textwidth]{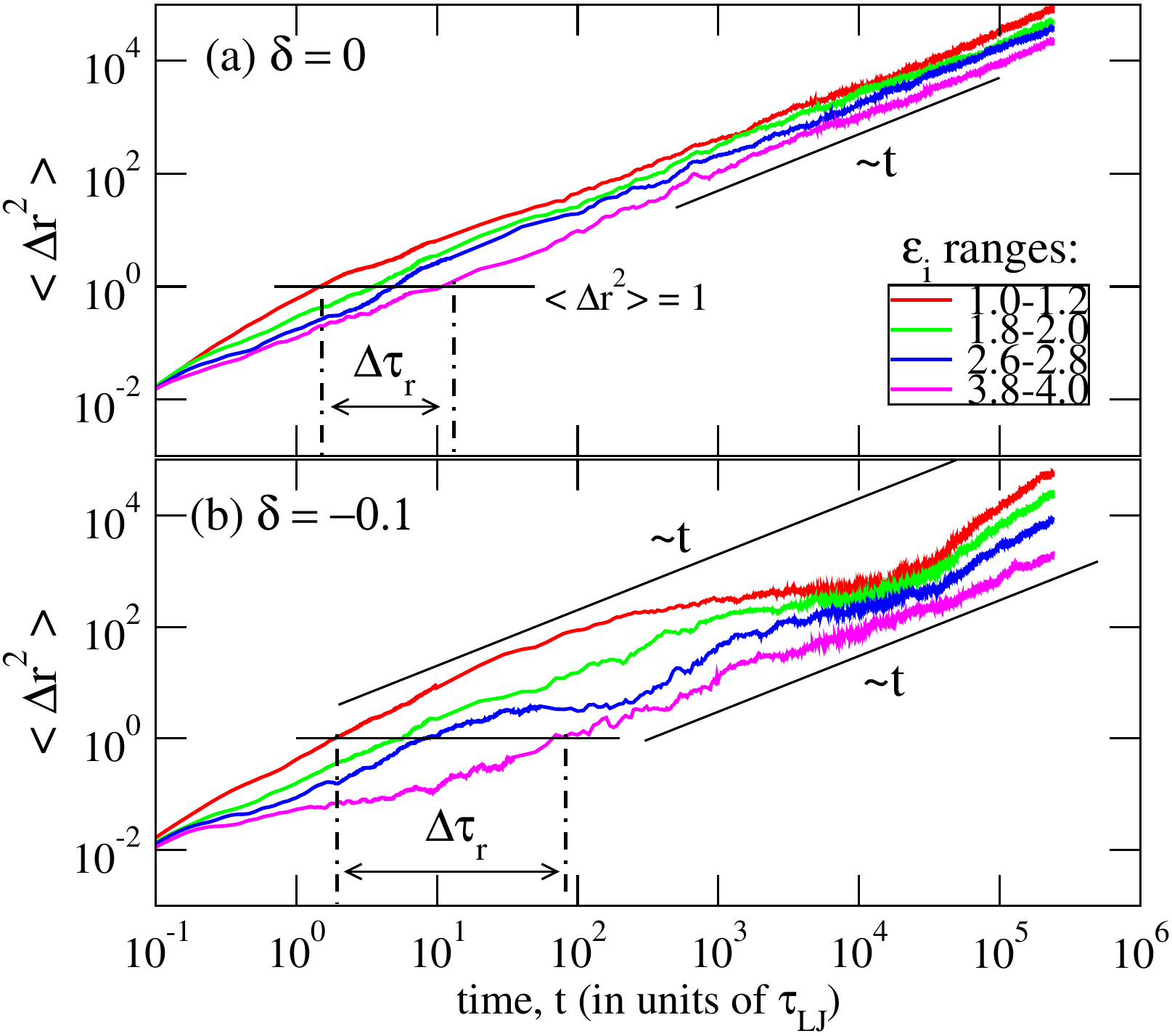}
\begin{center}
\caption{MSD for GM particles in four different $\epsilon_i$ ranges (as indicated in the upper panel),  at temperatures (a) $\delta=0$, and (b) $\delta=-0.1$. The intersection of the horizontal line at $\left\langle \Delta r^2 \right\rangle = 1$ (particle size) with the MSD curves of particles with the largest and the smallest $\epsilon_i$  defines the range $\Delta \tau_r$ (see text).}
\label{fig: msd-gm-epsi}
\end{center}
\end{figure}

In order to distinguish between MSDs of particles with different values of $\epsilon_i$,  we consider four narrow
($\Delta \epsilon_i = 0.2$) intervals of $\epsilon_i$ and plot the MSD as a function of time for these groups of particles
(we average over trajectories of particles in each group).
As shown in figure~\ref{fig: msd-gm-epsi}(a), at the transition $\delta=0$ (and at higher temperatures - not shown),
the four groups of particles exhibit normal diffusion with diffusion
coefficients that decrease with increasing interaction strength.
If we define a relaxation time $\tau_r$ as the time it takes to move one particle diameter on the average, we find that different types of particles have  relaxation times distributed over the
range $\Delta \tau_r$ with $\tau_r \sim 1$ for smallest $\epsilon_i$ particles
and $\tau_r \sim 10$ for the largest $\epsilon_i$ ones, as indicated in the figure.

As temperature is decreased into the supercooled fluid range $\delta\approx -0.1$ (fig.~\ref{fig: msd-gm-epsi}(b)),
the dynamics of the GM system becomes heterogeneous in the sense that the similarity between the MSD curves of particles with different values of $\epsilon_i$ breaks down and  $\Delta \tau_r$  becomes much broader due to dramatic increase of the relaxation time for largest
$\epsilon_i$ particles ($\tau_r \sim 100$).\
In order to understand the origin of this dynamic heterogeneity recall that the transition temperature $T^\ast~(\approx 1.0)$
reflects an average (over the different
particles in the system) property of the GM system; there are particles with $\epsilon_i > 2.5$ ($\epsilon_i < 2.5$ )
for which the transition temperature of a system made of such particles only, would be higher (lower) than that.
For example, at $T=T^\ast$ a one-component system of particles with $\epsilon_i \rightarrow 1$ will not freeze
because the transition temperature of such a system is about $0.4$ (2.5 times lower that that of a 1C system with $\epsilon_{i}=2.5$).
Therefore, for $T \rightarrow T^\ast$ one expects to observe crystalline clusters of larger $\epsilon_i$
particles that coexist with a fluid of mobile smaller $\epsilon_i$ particles
(see figs.~\ref{fig: config-gm-epsi}(a) and (b)). While this simple physical picture is clearly oversimplified since there are small $\epsilon_i$ particles
that are trapped inside the solid clusters, 
it agrees with the observation of  microphase separation between different types of particles in our system. The dynamical consequences of this segregation  in the vicinity of $T^\ast$ are responsible for the fact that large $\epsilon_i$ particles diffuse
relatively slowly compared to small $\epsilon_i$ ones, an effect that is clearly observed
in fig.~\ref{fig: msd-gm-epsi}(b).

We now focus on the dynamics of smallest $\epsilon_i$ particles at $\delta \approx -0.1$ (upper curve in fig.~\ref{fig: msd-gm-epsi}(b)) 
which diffuse much faster and cover larger distances than other particles. Since the former particles
are preferentially localized at the boundaries of the dense clusters, they can
escape into the voids and move freely inside them. This free diffusion in a confined space may explain the observation of normal diffusion at short time scales followed by leveling out of the MSD due to the finite size of the voids at later times $10^2< t<10^4$.
Beyond $t\approx 10^4$ there is further increase of MSD that may be related to surface-tension driven coarsening of the voids on such long time scales. Clearly, the dynamics of  the more mobile particles with the smallest values of $\epsilon_i$ is also responsible for the increase of particle-averaged MSD beyond $t\approx 10^4$ observed in fig.~\ref{fig: msd-compare}(b) (this mobile fraction gives the dominant contribution to large-scale displacements observed at very long times). If we now consider the dynamics of particles with the largest $\epsilon_i$ in fig.~\ref{fig: msd-gm-epsi}(b) we notice that their MSD undergoes a transition from a plateau at short time scales ($t<10-100$) to simple diffusion at longer times ($t>100$), as they escape from a localized state characterized by lattice vibrations on length scales smaller than the particle diameter.  For particles with intermediate $\epsilon_i$ values
the behavior of MSD is in-between the two extremes discussed above. \\

The above analysis gives us an crude picture of particle motions in our system. In order to obtain deeper insights
about the dynamics of the APD system, we now proceed to use other statistical methods of characterization 
of particle dynamics. One such method involves calculating the
distribution of relative angles between successive steps
along the trajectory \cite{SBurov2013_PNAS}. The sensitivity of this method has been demonstrated
in the study of caging and escape in a quasi-2d colloidal suspension, near the
jamming transition; in this regime the MSD curve is linear, while the angular
distribution shows a peak at $\theta=\pi$ (see fig.~5 in ref.~\cite{SBurov2013_PNAS}). 
We proceed to calculate the angle distributions for 1C and GM systems at different temperatures near the solidification transition.


\section{Angle distribution}
\label{sec: angular-dis}

For a given particle trajectory, we define a vector ${\bf V}(t;\Delta) = {\bf r}(t+\Delta) - {\bf r}(t)$, where ${\bf r}(t)$ is the position of the particle at time $t$ and $\Delta$ is the time lag. Then, the relative angle $\theta (t;\Delta)$ between two consecutive vectors ${\bf V}(t;\Delta)$ and ${\bf V}(t+\Delta; \Delta)$ along the trajectory is given by the dot product
\begin{eqnarray}
 \cos\theta(t;\Delta) = \frac{{\bf V}(t;\Delta)\cdot {\bf V}(t+\Delta; \Delta)}{\left| {\bf V}(t;\Delta) \right| \left| {\bf V}(t+\Delta; \Delta) \right|}~,
\label{eq: relative-angle}
\end{eqnarray}

\noindent where $\left| {\bf V}\right|$ is the magnitude of {\bf V}.
From eqn (\ref{eq: relative-angle}) we get $\theta (t;\Delta)$, and obtain
the distribution of angles $P(\theta;\Delta)$ by computing all the relative angles along
each trajectory and averaging over the trajectories. A peak at $\theta=0~{\rm or}~2\pi$ corresponds to correlated/directed
steps, whereas a peak at $\theta=\pi$ represents anti-correlated steps, e.g. due to
reflection from confining walls. By varying $\Delta$ we can examine the dynamics at
various levels of temporal coarse-graining. The upper bound for $\Delta$ is chosen
such that $\Delta < t^{\ast\ast}$, with $t^{\ast\ast}$ the time to diffuse across the system size. In the following, we first examine the angle distributions obtained by considering the trajectories of all the particles, and then compare the angle distributions of different types of particles by averaging over the trajectories of particles in a given interval $\Delta \epsilon_i $ only.

\begin{figure}[ht]
\includegraphics*[width=0.4\textwidth]{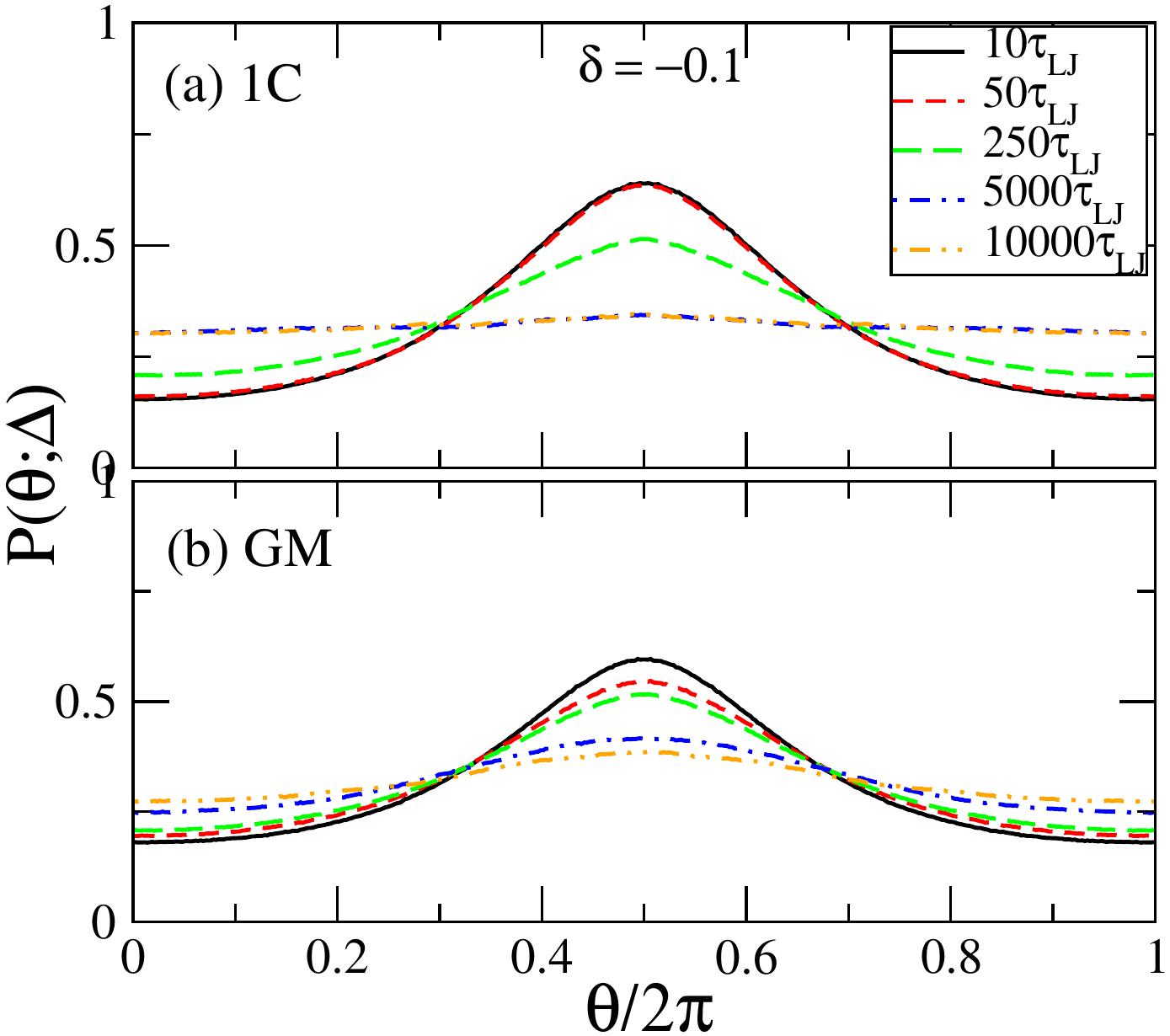}
\begin{center}
\caption{Angle distributions $P(\theta;\Delta)$ of (a) 1C and (b) GM systems obtained by
considering trajectories of all the particles at $\delta=-0.1$. Different line types
correspond to different values of temporal coarsening $\Delta$ as indicated in the inset.}
\label{fig: angular-dis-compare}
\end{center}
\end{figure}

 At high temperatures $\delta > 0$,
the distributions are uniform for both the 1C and the GM systems (not shown). As temperature is lowered below $T^*$, one expects to observe the effect of
confinement by neighboring particles (caging) at short time scales, followed by
 escape from the cage and diffusion at a longer times.
As shown in figure~\ref{fig: angular-dis-compare} for $\delta\approx -0.1$ and small values of  $\Delta$, both 1C and GM systems exhibit a peak
at $\theta=\pi$ in the angle distribution $P(\theta;\Delta)$ obtained by
averaging over the trajectories of all the particles.
With increasing $\Delta$ the height of the peak decreases and eventually the angle distributions become isotropic (i.e., uniform in the range $0-2\pi$), typical of normal diffusion where all the possible angles are equally probable.
The difference in the dynamics of the two systems becomes evident when one examines the rate of change of  $P(\theta;\Delta)$ upon increasing $\Delta$; the decrease of peak height is much more gradual in the case of GM and even at
$\Delta=10^4$ the distribution is not completely flat.
In the following, we will focus on the GM system and compare the behavior
of $P(\theta;\Delta)$ for different particle types.\

\begin{figure}[ht]
\includegraphics*[width=0.45\textwidth]{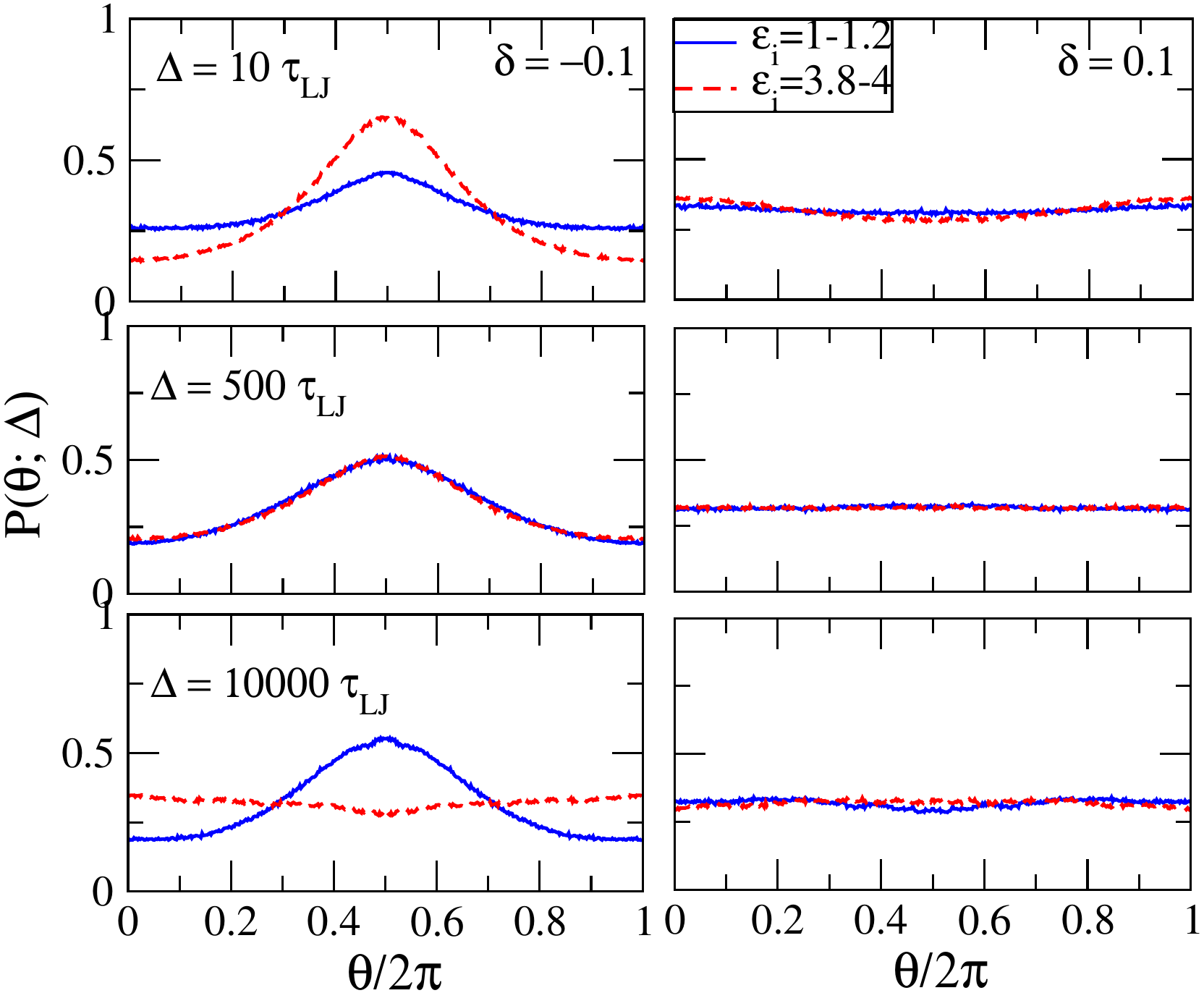}
\begin{center}
\caption{Comparison of angle distributions of the smallest and largest $\epsilon_i$
particles, shown for different $\Delta$ (top to bottom panel), at reduced temperatures $\delta\approx -0.1$ (left panels) and $\delta \approx 0.1$ (right panels).}
\label{fig: angular-dis-gm}
\end{center}
\end{figure}

The angle distributions of two different groups of particles
with $\epsilon_i$ in the ranges 1-1.2 and 3.8-4 respectively, are compared in figure~\ref{fig: angular-dis-gm}
for three different values of $\Delta$, at temperatures below and
above the liquid-solid transition.
At $\delta\approx -0.1$ and $\Delta=10$, a strong peak at $\theta=\pi$ is observed
for large $\epsilon_i$ particles, as expected for
particles confined within solid clusters. This peak decays slowly with increasing temporal coarsening
and disappears (actually, it is replaced by a shallow dip) at $\Delta=10^4$,
indicative of slow cage relaxation.
On the other hand, for small $\epsilon_i$
particles the confinement effect increases in strength at intermediate time scales
and remains strong up to the largest time lag
in our simulaton. Since such particles are expected to reside
preferentially at the periphery of solid clusters and inside voids,
this persistent behavior may be attributed to particles moving inside voids of different sizes
which are forming and breaking as time goes on, with the largest holes
being $\sim 30\sigma$ in diameter (see fig.~\ref{fig: config-gm-epsi} and supplementary  figs.~\ref{fig: s1}).
Notice that from the MSD of the largest $\epsilon_i$
particles in fig.~\ref{fig: msd-gm-epsi}(b), one might conclude that for
$t > 10$ the particles are already well into the normal diffusion regime,
but this is not quite correct since the angle distribution shows that strong confinement effects persist up to
$\Delta\sim 500$. This  demonstrates that
$P(\theta;\Delta)$ is a sensitive probe for dynamics which provides information
that could not be obtained from the MSD alone.\

At higher temperature ($\delta\approx 0.1$) both particle types
have uniform distributions in the range $0-2\pi$; however, it is important to note
that if we look very closely at the distributions for $\Delta=10~{\rm and}~10^4$, we realize that the curves 
do not exactly fall on top of each other. For instance, while at $\Delta=10$ the angle distribution is completely flat for small $\epsilon_i$ particles,
there is a dip at $\theta=\pi$ for large $\epsilon_i$ particles, indicating very
weakly correlated motion. Such weak correlations can not be detected by analyzing the MSD curves,
since all particle types have $\left\langle \Delta r^2 (t) \right\rangle \sim t$
when $t\sim 10$, even at $\delta\approx 0$ (see fig.~\ref{fig: msd-gm-epsi}(a)).\\

Although angle distributions provide information beyond the MSD, they involve
averaging over each trajectory (for fixed time intervals) and since the change of angle along
the trajectory may be same for particles exploring large and small regions of
space, some of the spatial information is lost. In order to provide detailed spatial and temporal information about the trajectories, we proceed to
study dynamic heterogeneities in the GM system using the displacement distribution.


\section{Displacement distribution}
\label{sec: displacements-distribution}

Distinct dynamical regimes are often observed in systems close to criticality. The
qualitative change in the dynamics across the transition is well captured by the
displacement distribution (van Hove self-correlation function) defined as
\begin{equation}
 G_s(r,\Delta) = \langle \delta(r-\lvert {\bf r}_i(\Delta) - {\bf r}_i(0) \lvert) \rangle~,
 \label{eq: van-hove-self}
\end{equation}
where $\Delta$ is the time interval, and ${\bf r}_i(\Delta)$ is the position of $i^{\rm th}$
particle at time $\Delta$. Eqn~(\ref{eq: van-hove-self}) describes the probability of
finding a particle at distance $r$ from a given initial position after time interval $\Delta$.

\begin{figure}[ht]
\includegraphics*[width=0.45\textwidth]{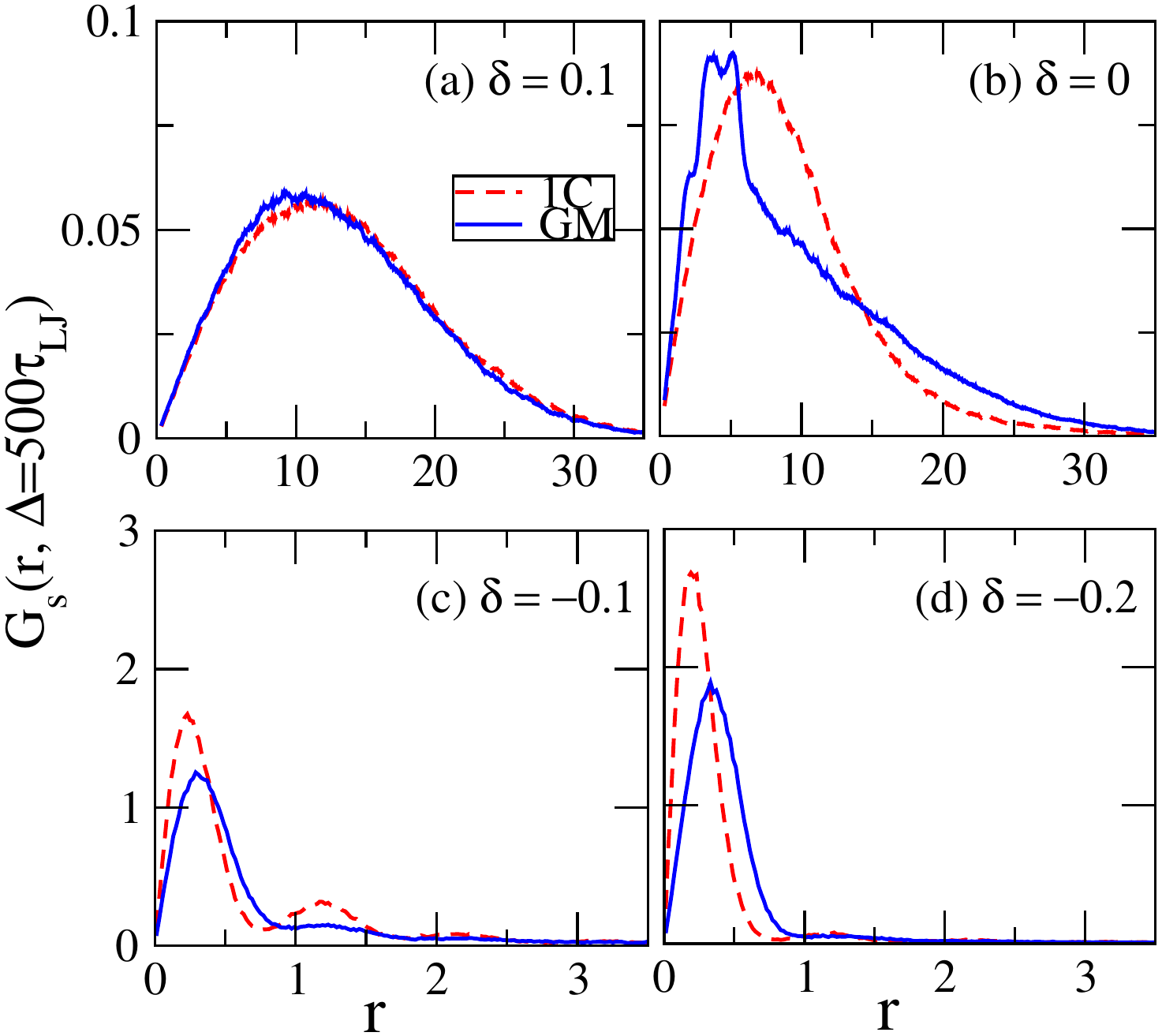}
\begin{center}
\caption{Comparison of the distribution of displacements at $\Delta=500$ for 1C and GM systems at
four different values of $\delta$ indicated in the figure.}
\label{fig: van-hove-self-part-compare-T}
\end{center}
\end{figure}

\begin{figure}[ht]
\centering
\includegraphics*[width=0.45\textwidth]{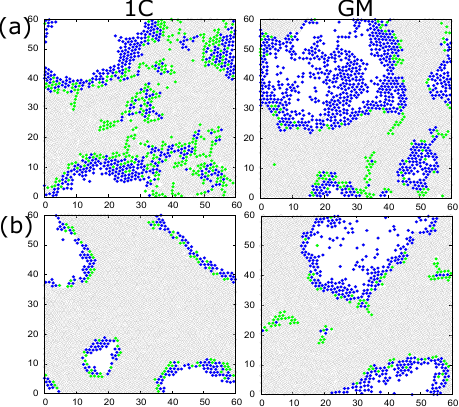}
\begin{center}
\caption{Typical configurations of 1C and GM systems with particles colored according to magnitude of displacement $r$ during time $\Delta=500$, shown for (a) $\delta\approx -0.1$ (upper panel), and (b) $\delta\approx -0.2$ (lower panel). Color code: gray open-circles for $r<0.75$, green filled-circles for $0.75\leq r \leq 1.5$, and blue filled-circles for $r > 1.5$.}
\label{fig: config-color-dr}
\end{center}
\end{figure}

In fig.~\ref{fig: van-hove-self-part-compare-T} we compare $G_s(r,\Delta)$ of 1C and GM systems at different reduced temperatures $\delta$, for $\Delta=500$. The distributions for different values of $\Delta$ at fixed $\delta$ ($\approx -0.2$) are shown in the SI fig.~\ref{fig: s2}.  At $T>T^\ast$ ($\delta=0.1$) the displacement distribution is Gaussian (since the normalization is with respect to the measure $dr$, the Gaussian is multiplied by $r$), as expected for simple diffusion (see fig.~\ref{fig: van-hove-self-part-compare-T}(a)).
A qualitative difference between the the displacement distributions of two systems is observed at $\delta\approx 0$ (see fig.~\ref{fig: van-hove-self-part-compare-T}(b)). Here, the displacement distribution decays exponentially for 1C and is a complicated non-exponential function for GM, (see also supplementary fig.~\ref{fig: s3}(a)).
For $\delta < 0$, the distributions have several peaks: a primary peak in the range $0< r \leq 0.75$, and a secondary peak in the range $0.75 \leq r \leq 1.5$, with the secondary peak being more pronounced for 1C system (see figs.~\ref{fig: van-hove-self-part-compare-T}(c)-(d)). 

\begin{figure}[ht]
\centering
\includegraphics*[width=0.5\textwidth]{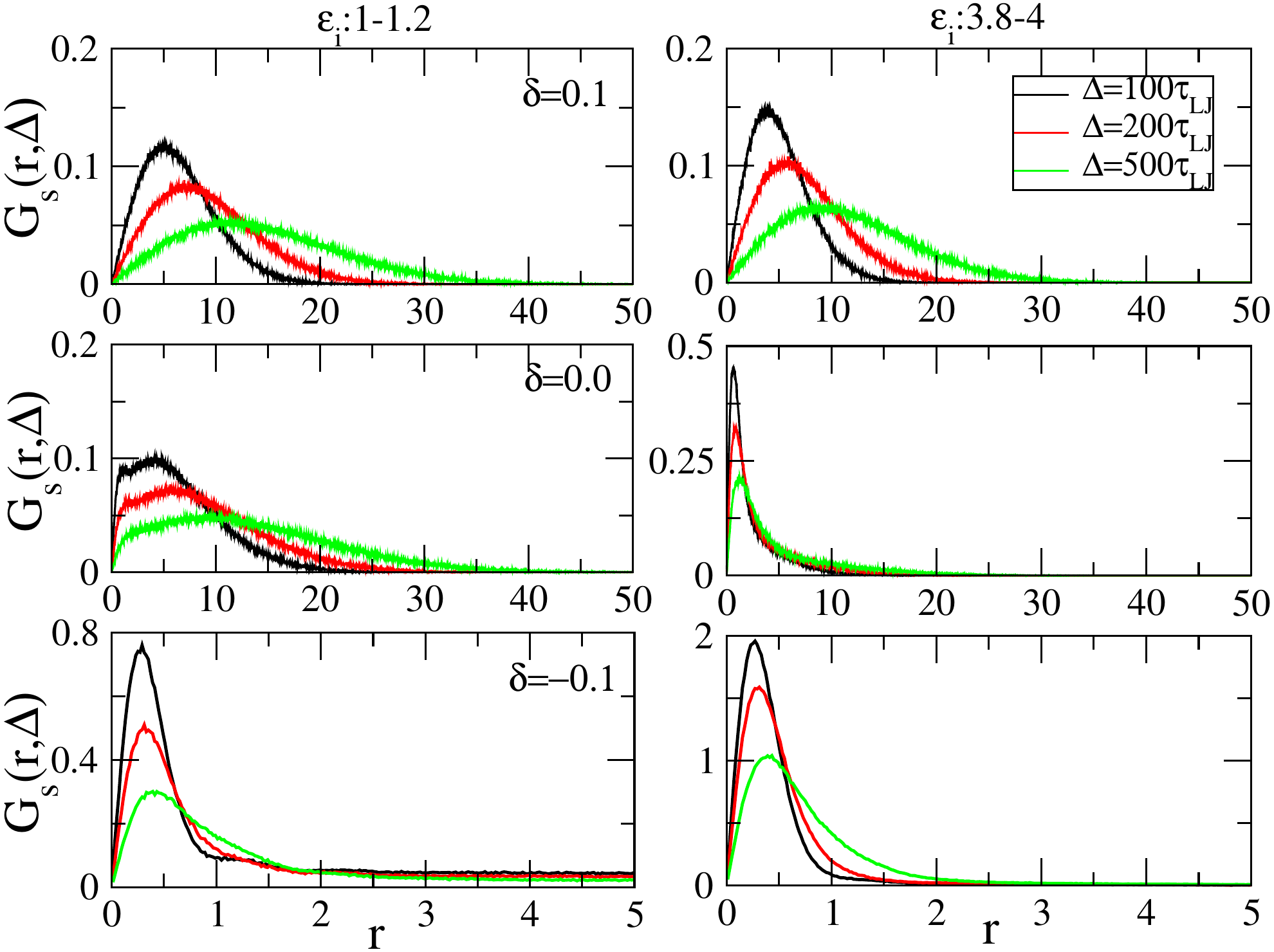}
\begin{center}
\caption{Displacement distributions of small and large $\epsilon_i$ particles (left and right panels, respectively) are shown for three different values of $\Delta$ (see inset), at three different temperatures $\delta$ (top to bottom panels).}
\label{fig: van-hove-gm-diff-ei}
\end{center}
\end{figure}

Insight into the origin of the multiple peaks of $G_s$ for $\delta<0$, can be obtained by coloring the particles according to the magnitude of displacement $r$. In fig.~\ref{fig: config-color-dr} we use three colors that correspond to three different ranges of displacement: (1) $r \leq 0.75$ (primary peak), (2) $0.75<r \leq 1.5$ (secondary peak), and (3) $r>1.5$, see figs.~\ref{fig: van-hove-self-part-compare-T} (c) and (d). Clearly, the first peak originates from vibrations of localized particles in solid clusters, the second peak arises from mobile particles associated with lattice defects and lastly, the higher order peaks that correspond to large displacements are due to mobile (fluid) particles located at the boundaries of the solid and inside voids. The relative change in the height of the peaks of $G_s$ with changing $\delta$ [compare figs.~\ref{fig: van-hove-self-part-compare-T} (c) and (d)] results from the corresponding change of the number of particles in different $r$ ranges in figs. \ref{fig: config-color-dr} (a) and (b).\\

So far we discussed the displacement distributions obtained by considering all the particles in the system. In the following, we compare the distributions of different particle types in GM system. As before, we consider two different groups of particles with $\epsilon_i$ in the ranges $1-1.2$ and $3.8-4$ respectively, and calculate the corresponding displacement distributions (fig.~\ref{fig: van-hove-gm-diff-ei}). Above the transition ($\delta\approx 0.1$) both particle types have qualitatively similar (Gaussian) distributions; however, the range is slightly larger for small $\epsilon_i$ particles. The distributions for small and large $\epsilon_i$ particles are qualitatively different at the transition ($\delta\approx 0$): $G_s(r,\Delta)$ is much narrower and decays more rapidly with $r$ for large compared to small $\epsilon_i$ particles. In the latter case the form of the distribution is quite complex, especially at shorter times, suggesting that small $\epsilon_i$ particles sample a broader range of environments. It is interesting to note that the displacement distributions of small and large $\epsilon_i$ particles at $\delta\approx 0$, resemble the distributions obtained in a single-molecule tracking study of two different types of nuclear proteins Dendra2 and H2B, respectively (see figure 2 of figure supplement 1 in reference~\cite{IIzeddin2014_eLife}). Indeed, the transcription factor Dendra2 behaves as a freely diffusing (and thus weakly interacting) particle, while H2B which is strongly bound to chromatin, displays a more restricted motion.
At $\delta\approx -0.1$, the fraction of caged particles  (with $r<$~particle diameter) increases dramatically for all particle types and, as expected, the amplitude of the effect is largest for large $\epsilon_i$ particles. At longer times (larger $\Delta$), the distributions broaden indicating that some particles are able to escape from their cages. The distribution decays much faster with $r$ for large than for small $\epsilon_i$ particles.\\


\section{Waiting-time distributions and cage remodeling time}
\label{sec: waiting-time-analysis}

\begin{figure}[ht]
\includegraphics*[width=0.47\textwidth]{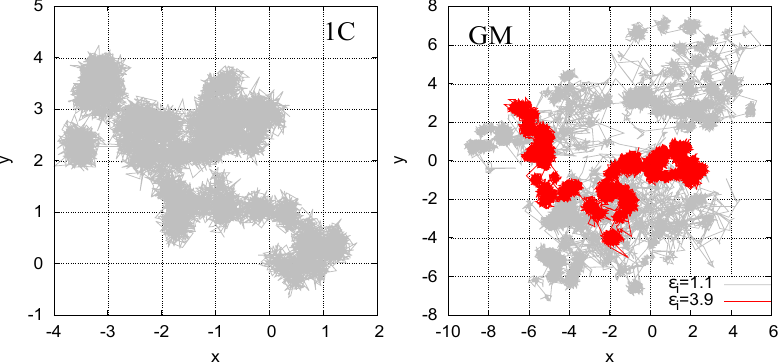}
\begin{center}
\caption{Typical trajectories of particles in the bulk solid phase for 1C and GM systems at $\delta\approx -0.2$.The two GM trajectories correspond to particles with $\epsilon_i=1.1$ and $3.9$, respectively. Particle positions are stored at intervals of $10$ and the trajectories are of length $2\times 10^5$ (in units of $\tau_{lj}$).}
\label{fig: trajectory-1c-gm-lowT}
\end{center}
\end{figure}

Inspection of particle trajectories at low temperatures shows that their  motion is intermittent, i.e. consists of long periods of localized vibrations followed by sudden jumps (see fig.~\ref{fig: trajectory-1c-gm-lowT}). Similar phenomena are observed in a wide variety of systems that range from relaxation in disordered materials to protein trajectories in live cells.\cite{Bouchaud1990_PhysRep,IIzeddin2014_eLife,AVWeigel2011_PNAS} Inspection of the trajectories shown in fig.~\ref{fig: trajectory-1c-gm-lowT} suggests that there is a broad distribution of trapping times (the amount of time particles spend in the cage, before hoping to another site). This leads to anomalous diffusion that can be described e.g., in the framework of the continuous time random walk (CTRW) model, in which a particle is trapped at a site for a random waiting time $\tau_w$ before jumping to the next site  (for a discussion of this and other models of anomalous diffusion, see ref.~\cite{RMetzler2014_PCCP}). In this model, the probability density function (pdf) of $\tau_w$ follows the power-law form
\begin{equation}
P(\tau_w) \sim \tau_w^{-(1+\alpha)}~,
\label{eq: waiting-time}
\end{equation}
with $0<\alpha<1$.  Since the CTRW model has a distribution of waiting times and jump lengths, mapping an MD trajectory on a CTRW is non-trivial and requires filtering the continuous trajectories obtained from MD simulations \cite{JHelfferich2015_PRE1}.  However, in the present study we follow a simpler approach and calculate the time a particle remains confined within a certain threshold radius ${\rm R_{th}}$ \cite{AVWeigel2011_PNAS}.  
Since $\tau_w$ depends on ${\rm R_{th}}$ one needs to check whether the exponent $\alpha$ is affected by the choice of ${\rm R_{th}}$. In the following we will take ${\rm R_{th}}=1$ which is about the mean inter-particle separation in the solid phase and thus corresponds to the typical cage size. Therefore, $P(\tau_w)$ represents the distribution of dwell times of particles in a cage (the distribution $P(\tau_w)$ for ${\rm R_{th}}=5$ is shown in the supplementary fig.~\ref{fig: s4}).\

\begin{figure}[ht]
\includegraphics*[width=0.45\textwidth]{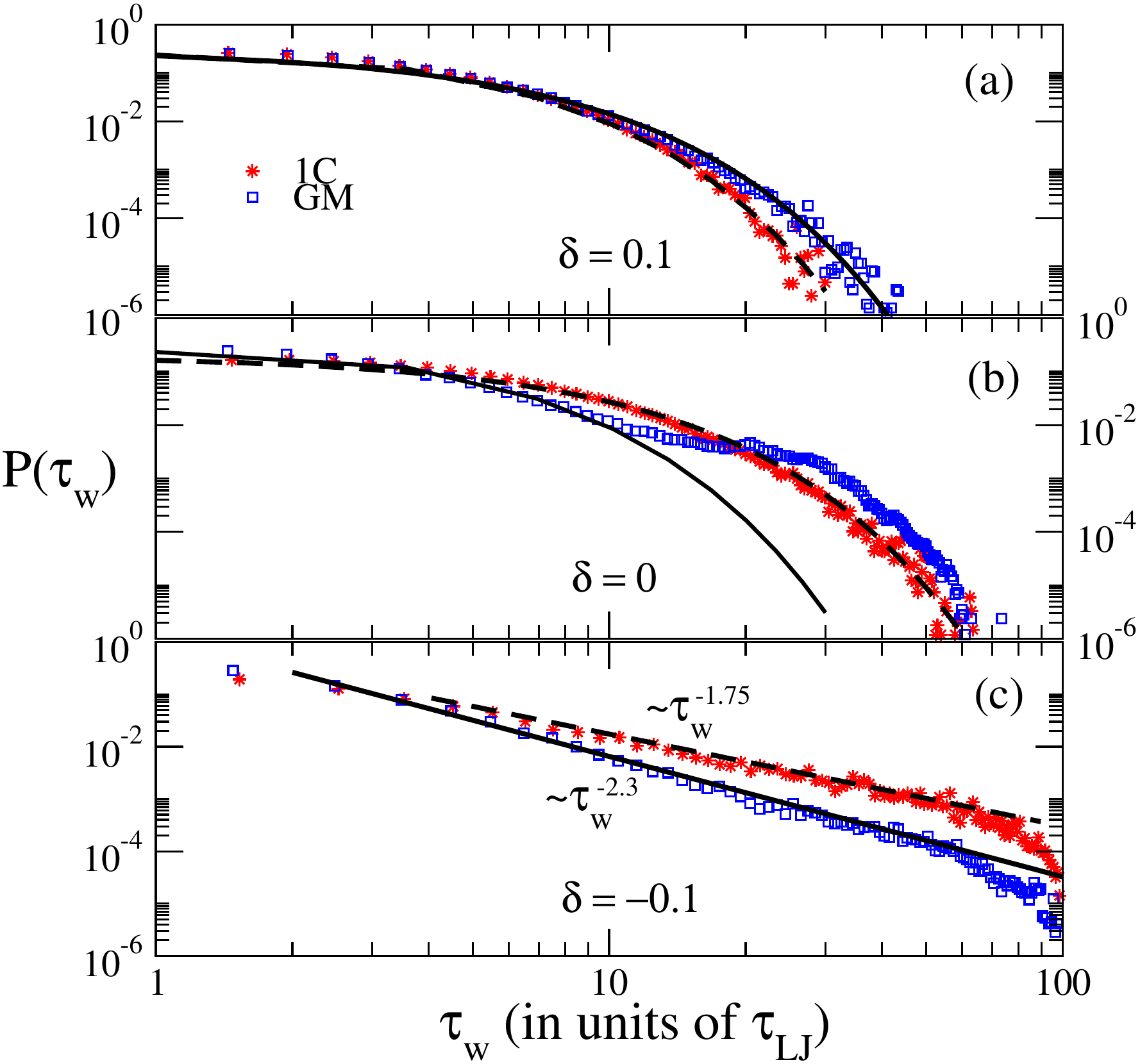}
\begin{center}
\caption{Waiting time distributions $P(\tau_w)$ (obtained by considering all particles)
at three different values of $\delta$, for $R_{\rm th}=1$. Exponential fits are shown by solid ($\tau = 2.35$) and dashed ($\tau = 2.25$) lines in (a) and by solid ($\tau = 2.35$) and dashed ($\tau = 5$) lines in (b), for GM and 1C systems, respectively.  In fig.~(c) the solid and dashed lines are power-law fits.}
\label{fig: waitingtime-dis}
\end{center}
\end{figure}

 For normal diffusion the waiting-time pdf corresponds to a  Poisson process and is therefore exponential,
\begin{equation}
P(\tau_w) = \tau^{-1}\exp(-\tau_w/\tau)~,
\label{eq: possionian-waitingtime}
\end{equation}
with $\tau$ the variance, see fig.~\ref{fig: waitingtime-dis}(a). At $\delta\approx 0$, there is a qualitative difference between the waiting time distributions of the two systems: while $P(\tau_w)$ of the 1C system is still exponential, that of the GM system can be fitted by an exponential only at short times, $\tau_w<10$ (fig.~\ref{fig: waitingtime-dis}(b)).  Note that while in this range of waiting times, the pdf of the GM system decays faster with $\tau_w$ than that of the 1C system, the situation is reversed at longer time scales. For $\delta<0$, the waiting time distribution is well-fitted by a power-law [eq.~(\ref{eq: waiting-time})] with $\alpha\approx 0.75$ for 1C system and $\alpha\approx 1.3$ for GM system (fig.~\ref{fig: waitingtime-dis}(c)), indicating that in this temperature range escape from a cage is easier, on average, for particles in the GM system than in the 1C system.

\begin{figure}[ht]
\includegraphics*[width=0.45\textwidth]{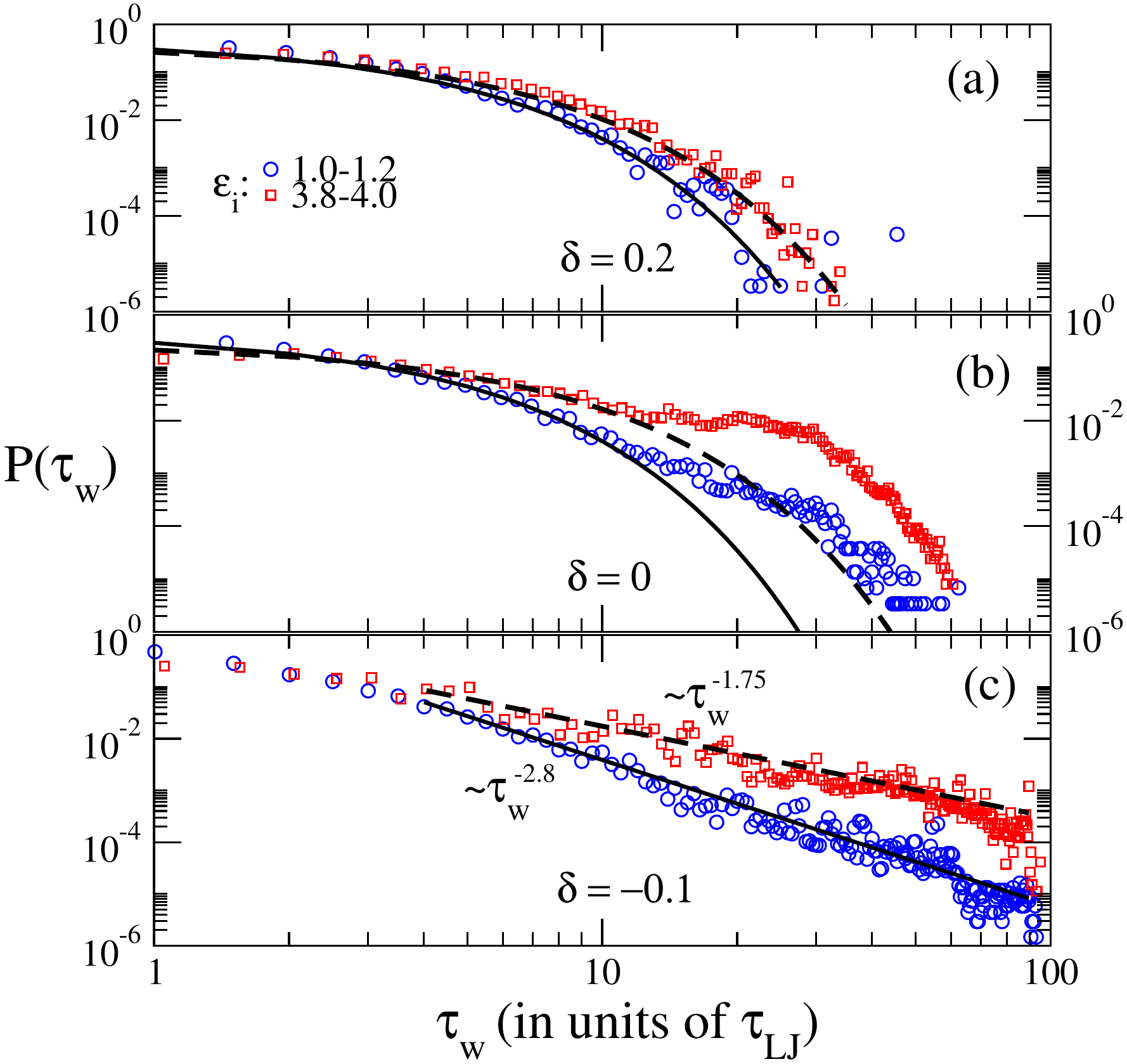}
\begin{center}
\caption{$P(\tau_w)$ compared for two different types of particles in GM system. Solid and dashed lines in (a) and (b) are
exponential fits to the distributions for $\epsilon_i:$ 1-1.2 and $\epsilon_i:$ 3.8-4, respectively. The corresponding decay times
are (a) 2.1 (solid) and 2.8 (dashed), (b) 2.1 (solid) and 3.5 (dashed). 
(c)  power-law fits.}
\label{fig: waitingtime-dis-epsi-banding}
\end{center}
\end{figure}

\begin{figure}[ht]
\includegraphics*[width=0.45\textwidth]{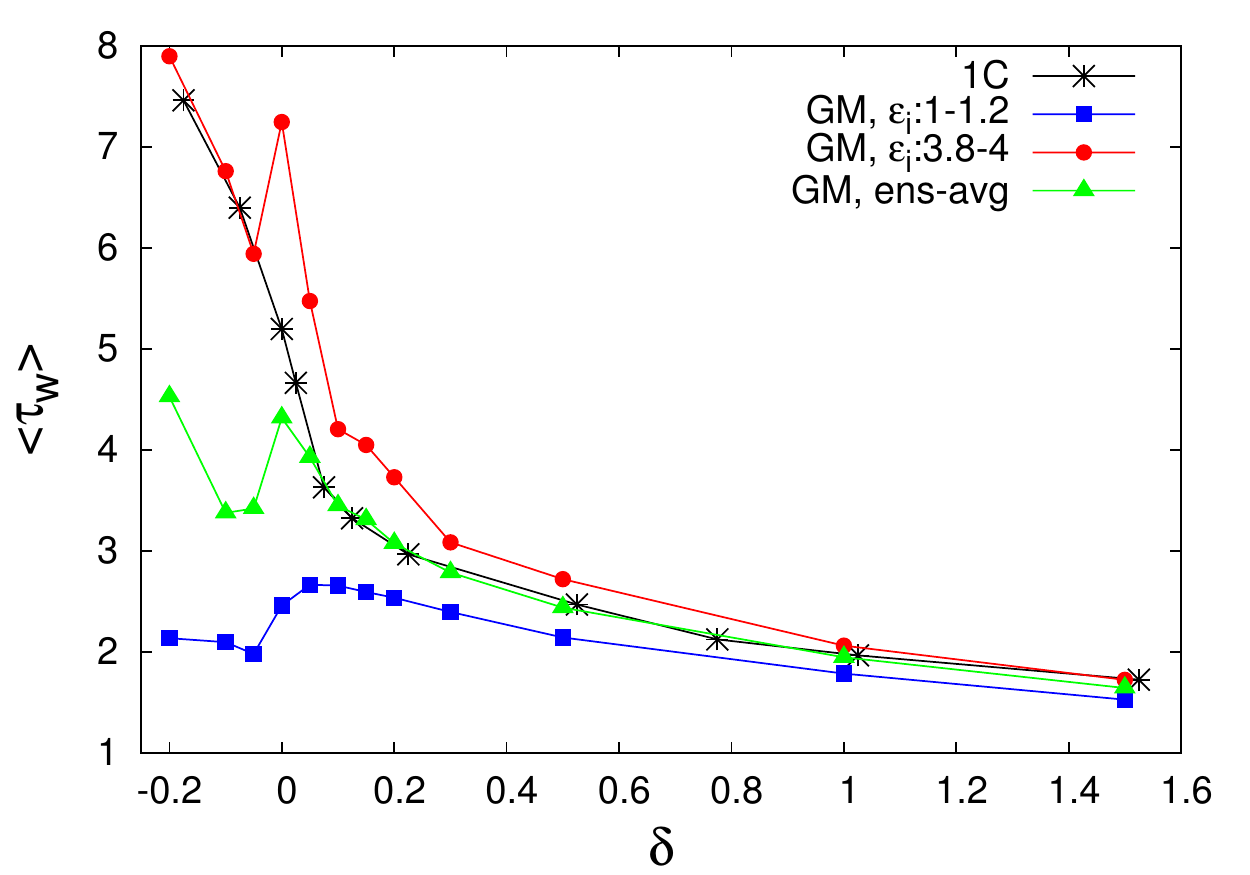}
\begin{center}
\caption{Mean waiting time $\left\langle \tau_w \right\rangle$ as a function of $\delta$ for 1C and GM systems. With increasing $\delta~(>0)$ the value of $\left\langle \tau_w \right\rangle$ converges to the same value, irrespective of particle type.}
\label{fig: wtime-mean-compare}
\end{center}
\end{figure}

An interesting question is whether different particle types in GM system have similar or distinct waiting time distributions. To this end we computed $P(\tau_w)$ for smallest and largest $\epsilon_i$ particles and compared them at different temperatures, see fig.~\ref{fig: waitingtime-dis-epsi-banding}. As one would expect, at higher temperatures $\delta>0$, escape from the cage is always a Poisson process and $P(\tau_w)$ is well-fitted by an exponential distribution, eqn~(\ref{eq: possionian-waitingtime}), though the relaxation time is shorter for low $\epsilon_i$ particles. At the transition $\delta\approx 0$, the curves have a more complex form and a broad transition region from exponential to power law behavior (with a plateau in the range $10<\tau_w<30$) is observed in fig.~\ref{fig: waitingtime-dis-epsi-banding}(b). At $\delta<0$ both curves are well-fitted by a power law $P(\tau_w) \sim \tau_w^{-(1+\alpha)}$, see fig.~\ref{fig: waitingtime-dis-epsi-banding}(c). The fits yield $\alpha\approx 0.75$ for large $\epsilon_i$ and $\alpha\approx 1.8$ for small $\epsilon_i$ and the mean of these two values roughly corresponds to the exponent $\alpha\approx 1.3$ observed for GM system where we did not distinguish between particle types, see fig.~\ref{fig: waitingtime-dis}(c). Comparison of figs. fig.~\ref{fig: waitingtime-dis}(c) and \ref{fig: waitingtime-dis-epsi-banding}(c) shows that in the low temperature range, the power-law exponent corresponding to high $\epsilon_i$ particles in the GM system is the same as that of the 1C system, suggesting that the dynamics of escape from a cage in a crystalline solid is dominated by (universal) excluded volume repulsions and does not depend on the strength of attractive interactions. 

A plot of the mean waiting time $\left\langle \tau_w \right\rangle$ as a function of the reduced temperature $\delta$ is shown fig.~\ref{fig: wtime-mean-compare}. Note that for $\delta>0$  the value of $\left\langle \tau_w \right\rangle$ for the 1C system matches that of the particle-averaged GM system but for $\delta < 0$ the average waiting time for the 1C system coincides with that of the highest $\epsilon_i$ GM particles. The above results concur with the previously made observation that the behavior of a particle in its effective cage is dominated by geometric (steric repulsion) effects that are similar for all the particles and depends only weakly on the attractive forces that distinguish between them. Thus, at high temperatures, the system is in a homogeneous liquid state and the effective cage is very similar for the 1C and the GM particles. Below the transition, most particles in the 1C system and the higher $\epsilon_i$  particles in the GM system are in a crystalline (hexagonally ordered) state and again experience very similar ``cages''. Below the transition, GM particles with small $\epsilon_i$ values find themselves inside and at the boundaries of voids and are therefore less constrained than in the liquid state, resulting in low, nearly temperature-independent values of $\left\langle \tau_w \right\rangle$. The origin of the sharp peak of $\left\langle \tau_w \right\rangle$ at $\delta\approx 0$ for GM system is unclear at present but it is clearly associated with the transition between the liquid and the solid cages that takes place at this temperature. 

\begin{figure}[ht]
\includegraphics*[width=0.48\textwidth]{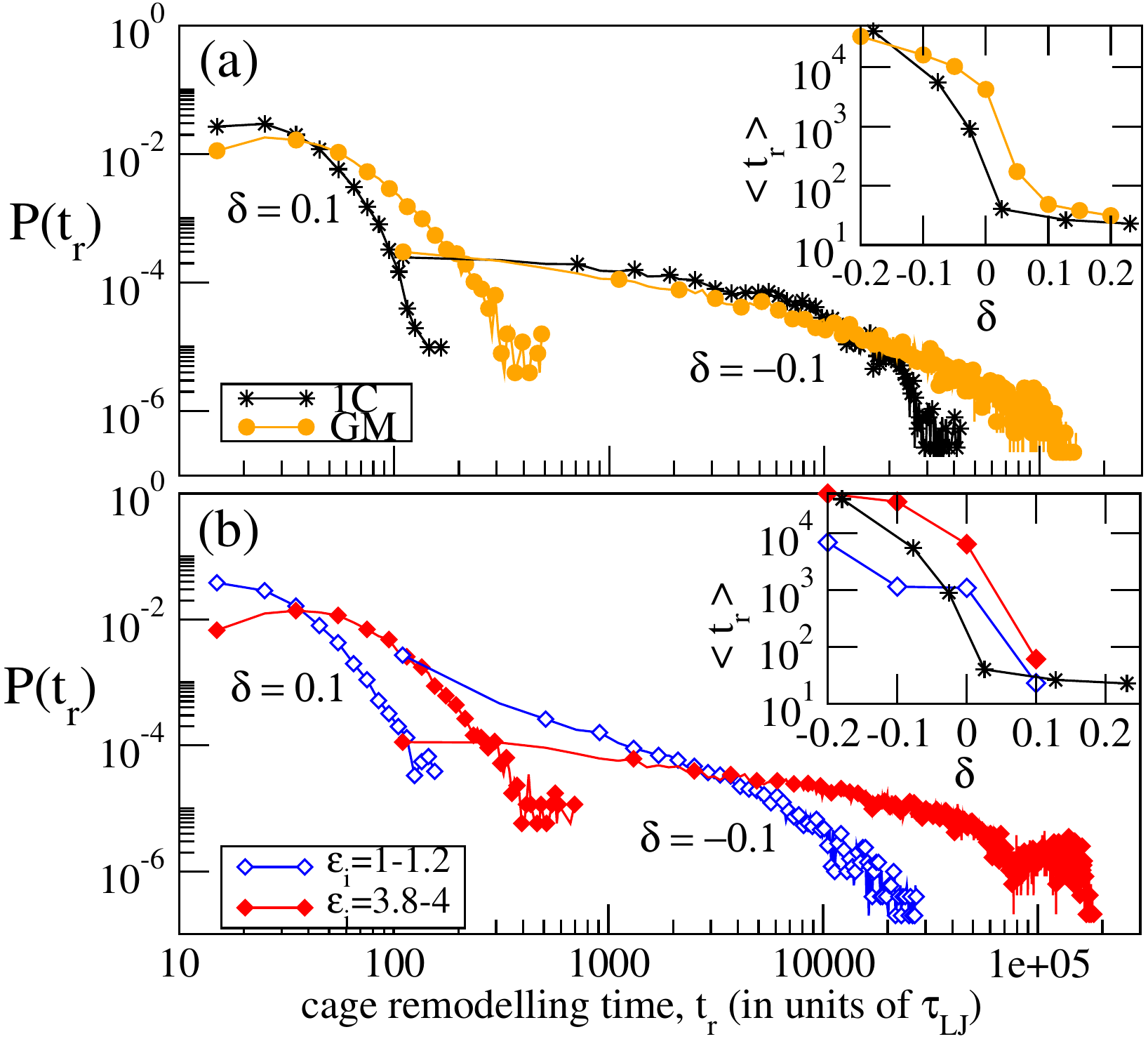}
\begin{center}
\caption{ Distribution of cage remodeling times $P(t_r)$ at two different values of $\delta$ indicated in the figure: (a) for 1C and GM (all particles are included) systems. (b) $P(t_r)$ for highest and lowest $\epsilon_i$ particles. The corresponding mean remodeling times are shown in the insets (the mean remodeling time for 1C is shown for comparison in the inset of (b)).}
\label{fig: cage-remodel-compare}
\end{center}
\end{figure}

Finally, we consider the distribution of cage remodeling times $P(t_r)$ where cage remodeling time $t_r$ is defined as the time to completely change the identity of the cage formed by the nearest-neighbors of a reference particle (note that this time is much longer than the escape time from a cage). In figure~\ref{fig: cage-remodel-compare}(a) we compare $P(t_r)$ for 1C and GM systems, computed by considering all the particles in the system. For $\delta \rightarrow 0$, the distribution $P(t_r)$ decays faster for 1C system which thus has a smaller $\left\langle t_r \right\rangle$, see fig.~\ref{fig: cage-remodel-compare}(a). There is abrupt increase in the value of $\left\langle t_r \right\rangle$ at the onset of freezing $\delta \rightarrow 0_+$, see  inset in figure~\ref{fig: cage-remodel-compare}(a). Since particles in the GM system are different, the cage remodeling time depends on particle type: the pdf decays faster (slower) for particles with small (larger) $\epsilon_i$, see figure~\ref{fig: cage-remodel-compare}(b); consequently, as shown in the inset, the value of $\left\langle t_r \right\rangle$ is always higher for larger $\epsilon_i$ particles (at $\delta\approx -0.1$ the difference between $\left\langle t_r \right\rangle$ of smallest and largest $\epsilon_i$ particles is more than an order of magnitude). Therefore, one can view the GM system below the transition as a system of mobile low $\epsilon_i$  particles diffusing inside the voids formed by the slowly rearranging solid clusters made of larger $\epsilon_i$ particles.


\section{Summary and discussion}
\label{sec: conclusion}

We have studied the dynamics of particles in a multi-component fluid model where all the particles in the system are different in the sense that each particle has a different interaction strength $\epsilon_i$. The local self-organization of particles according to their identity becomes more pronounced as temperature is decreased throughout the fluid region, as shown in fig.~\ref{fig: eeff-mean-gm} (see also reference \cite{Shagolsem2015_JCP}). The only effect of this neighborhood identity ordering on the purely diffusive motion of individual particles above the liquid-solid transition temperature is that different particle types 
move with different diffusion coefficients. This dynamical ``degeneracy'' is lifted at and below the freezing transition as the coupling between long-range hexagonal order and local neighborhood identity ordering (ordered solid domains are enriched by high $\epsilon_i$ particles while low $\epsilon_i$ particles proliferate in and around the low density regions (voids) and at mobile lattice defects) leads to heterogeneous dynamical behavior. 

Major differences between the dynamics of one-component systems and systems in which all particles are different are observed even when one considers ensemble-averaged quantities. At high temperatures, analysis of  MSD shows that on average particle moves by simple diffusion and that the diffusion coefficients are nearly identical in one-component and APD systems (both exhibit Arrhenius-like behavior, with roughly the same activation energy). Below the freezing temperature, the MSD curve of the GM system is consistently above that of the 1C system, an effect that becomes more pronounced at late times due the presence of relatively large number of mobile particles inside voids in the GM system. Analysis of angle distributions reveals the effects of particle caging  in both systems, but the transition to uniform angle distribution (corresponding to normal diffusion) with increasing temporal coarse-graining is more gradual for GM systems. The displacement distributions of the two systems are qualitatively similar for $\delta < 0$ and $\delta > 0$, but  are strikingly different at the transition ($\delta \approx 0$), with much slower decay of spatial correlations in GM system. The waiting time distributions are quite similar for $\delta > 0$ and follow Poisson statistics, while for $\delta < 0$ both distributions have a power-law form but with different exponents: $\alpha\approx 0.75$ (1C) and $\alpha\approx 1.3$ (GM).\

The rich dynamics of the GM system is fully revealed when one groups together particles with similar interaction strengths ($\epsilon_i$) and analyzes the trajectories of groups of particles with low and high $\epsilon_i$ separately. Indeed, as has been demonstrated in protein tracking experiments in living cells~\cite{IIzeddin2014_eLife}, particles of different types moving in the same complex environment have very different dynamical histories, because of the different ways in which they interact with the surrounding medium. Thus, although  particles of all types undergo normal diffusion in the liquid regime $\delta \ge 0$, their diffusion coefficients $D$ depend on particle type: lower (higher) $D$ for larger (smaller) $\epsilon_i$ particles. This quantitative difference between the motions of different types of particles becomes qualitative at and below the transition, as revealed by the various methods of analysis: MSD [fig.~\ref{fig: msd-gm-epsi}(b)], angle [fig.~\ref{fig: angular-dis-gm}], displacement [fig.~\ref{fig: van-hove-gm-diff-ei} and waiting-time distributions [fig.~\ref{fig: waitingtime-dis-epsi-banding}(b)-(c)]. Even though the way different particles move along their trajectories cannot be described in terms of simple diffusion or sub-diffusion, using the above methods combined with snapshots of the particle configurations, we were able to construct a relatively simple physical picture of the  dynamical phenomena that take place in supercooled APD fluids below the liquid-solid transition. Particles with higher interaction strength $\epsilon_i$ form large hexagonally-ordered clusters that contain mobile lattice defects and voids of different sizes; particles with lower interaction strength are preferentially localized in and at the boundaries of these voids and at lattice defects. While on short time scales high interaction strength particles vibrate about their local equilibrium positions in the cage formed by their neighbors, at longer times they move around due to a combination of processes which include escape from the cage and rearrangement of the solid clusters due to the motion of defects and surface-tension driven coalescence of voids. Weakly interacting particles are enriched in the fluid phase at the boundaries of the solid and inside voids and therefore the ``cages'' in which they move reflect the distribution of void sizes. The separation of time scales allows us to propose a very crude but qualitatively correct picture of the heterogeneous dynamics of the APD system in the supercooled regime: strongly interacting particles determine the geometry of the space in which weakly interacting particles move, with the result that the latter particles diffuse in a crowded and confined environment, reminiscent of geometry-controlled kinetics in ref.  \cite{Benichou}. However, in our case this geometry changes slowly with time and it is this combination of different time scales that leads to the complex motions observed. 

At yet lower temperatures than those reported in the present study (around $\delta\approx -0.5$) both the one-component and the APD systems become kinetically frozen on simulation time scales, in the sense that all the particles have condensed into the crystalline phase and the defects are immobile. Note that since particles can no longer exchange their positions, NIO ordering is arrested as well and one obtains an ``APD glass''. As we have shown lately in a study of the random bond model with particle exchange on a lattice \cite {dino}, this temperature is close to the onset of the annealed to quenched transition below which different realizations of the quenched distribution of interaction parameters $P(\epsilon_{ij})$ yield different results.

\begin{acknowledgements}
Helpful discussions with  D. Osmanovic, D. Kessler, J. Schulz, E. Barkai, and S. Burov are
greatfully acknowledged. YR's research was supported by the I-CORE Program of
the Planning and Budgeting committee and the Israel Science Foundation grant 1902/12, and
by the US-Israel Binational Science Foundation grant 2010036.
\end{acknowledgements}


\clearpage

\newpage
\onecolumngrid
\renewcommand{\thepage}{S\arabic{page}}
\renewcommand{\thesection}{S\arabic{section}}
\renewcommand{\thetable}{S\arabic{table}}
\renewcommand{\thefigure}{S\arabic{figure}}
\renewcommand{\theequation}{S\arabic{equation}}
\setcounter{figure}{0}
\setcounter{equation}{0}
\setcounter{table}{0}
\setcounter{section}{0}
\newpage

\begin{center}
{\Large SUPPLEMENTARY INFORMATION}
\end{center}

\section{Temperature dependence of diffusion coefficient}
\label{lab: arrhenius-plot-of-D}

\begin{figure}[ht]
\includegraphics*[width=0.55\textwidth]{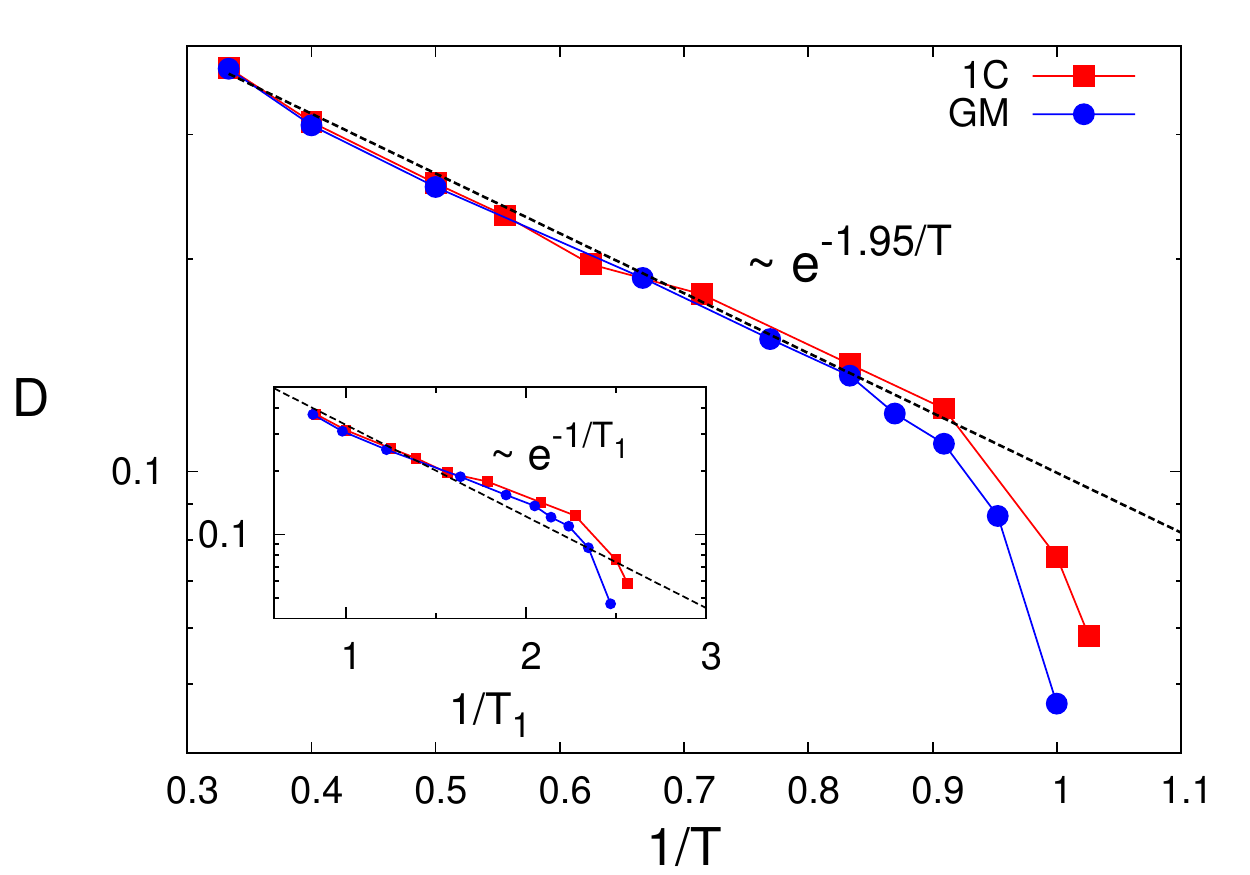}
\begin{center}
\caption{Diffusion coefficient $D$ vs $1/T$ (Log-linear plot). Temperature dependence of the diffusion coefficient $D(T)$. The inset shows the diffusion constant as a function of the rescaled temperature $T_1=T/<\epsilon_i^{\rm eff}(T)>$.}
\label{fig: diffusion-coeff-temperature}
\end{center}
\end{figure}

The temperature dependence of the diffusion coefficient is studied for $T\geq T^\ast$, see fig.~\ref{fig: diffusion-coeff-temperature}, where the diffusion coefficient $D$ is obtained from the ensemble average MSD. $D(T)$ shows Arrhenius-like behavior for large $T$: 
\begin{equation}
D(T) = D_0 e^{-E_a/(k_BT)}~,
\end{equation} 
with $D_0$ the diffusion coefficient at infinite temperature, $E_a$ the activation energy, and $k_B$ the Boltzmann constant. For $T\rightarrow T^\ast$, $D(T)$ deviates from the Arrhenius-like behavior indicating the onset of dynamics slowing down. From the fit in the Arrhenius regime we obtained $E_a$ and find that both 1C and GM systems have roughly the same value of activation energy $E_a\approx 1.95$. Similar values of $E_a$ for both the system is expected since the effective interaction parameter $\left\langle \epsilon_i^{\rm eff} \right\rangle$ which defines the local particle interaction is very close to that of the 1C system and weakly depend on $T$, see fig.~(3) in reference~\cite{Shagolsem2015_JCP}. \\

\newpage
\section{Supplementary figures}

\begin{figure}[ht]
\includegraphics*[width=0.5\textwidth]{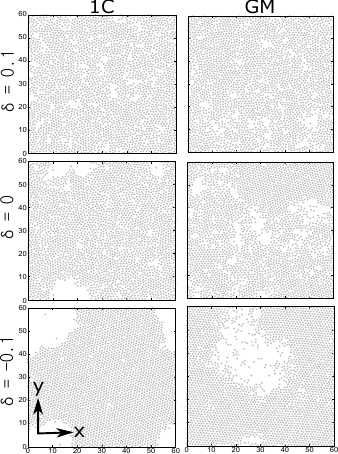}
\begin{center}
\caption{Typical configurations of the 1C and GM systems compared at the same distance from
the respective transition temperature. Shown in the figure for $\delta = (T-T^\ast)/T^\ast$ = 0.1, 0.0, and -0.1 (Top to bottom panel).}
\label{fig: s1}
\end{center}
\end{figure}

\begin{figure}[ht]
\includegraphics*[width=0.65\textwidth]{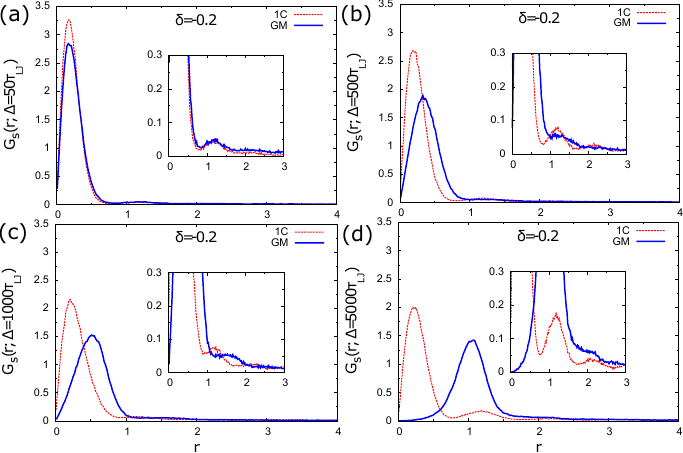}
\begin{center}
\caption{{\bf Distribution of displacements $G_s(r,\Delta)$:} The distributions for 1C and APD systems are compared at different values of lag-time $\Delta$ when $\delta\approx -0.2$. Inset is the zoom in showing secondary peaks. The presence of distinct dynamical regimes in the system is indicated by the secondary peaks. For small $\Delta$ the secondary peak is same for all the systems and it is enhanced upon increasing $\Delta$. On average the particles in GM system cover relatively greater distance.}
\label{fig: s2}
\end{center}
\end{figure}

\begin{figure}[ht]
\includegraphics*[width=0.65\textwidth]{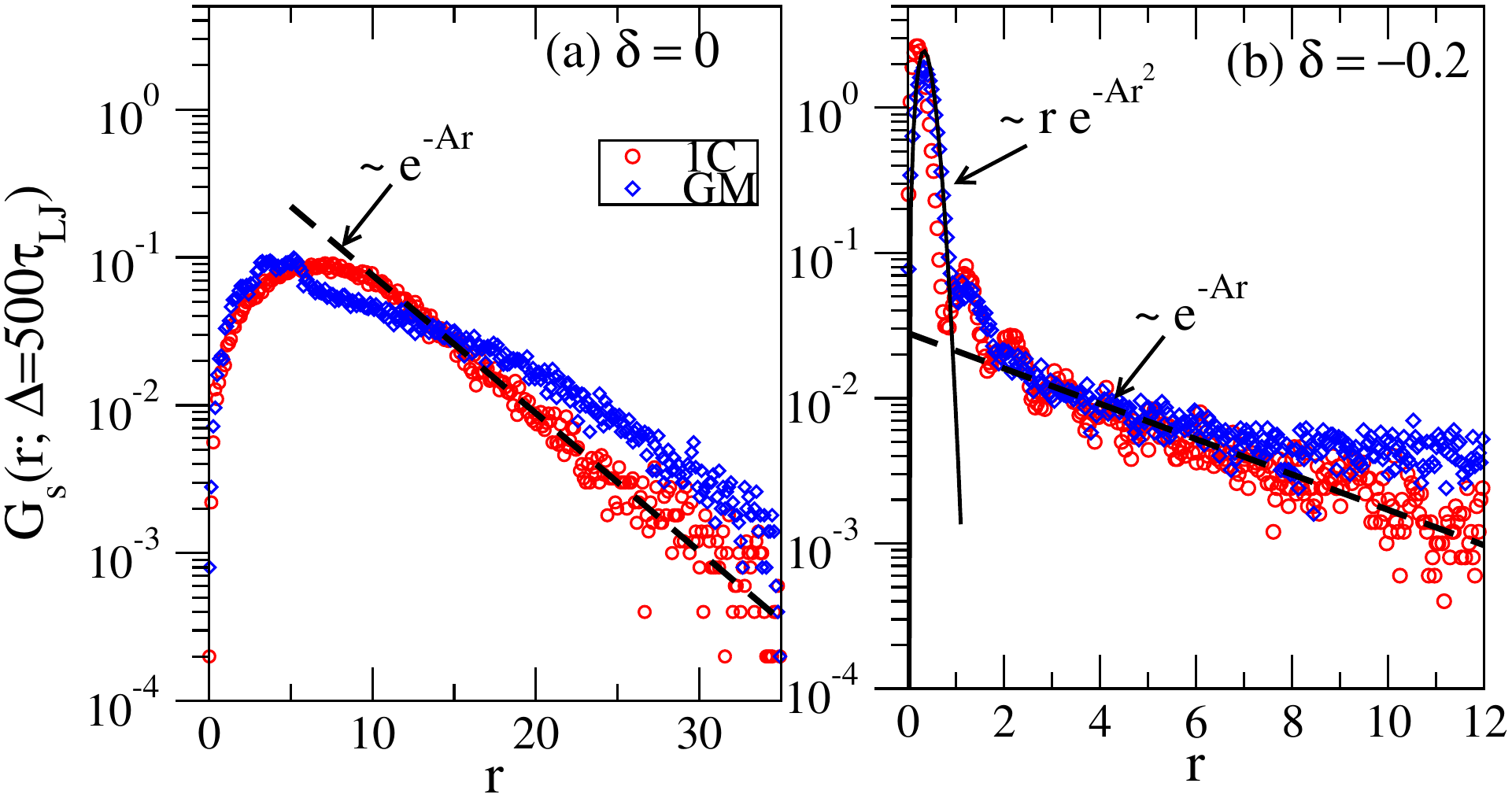}
\begin{center}
\caption{Log-linear plot of the displacements distributions shown in fig.~\ref{fig: van-hove-self-part-compare-T} for (a) $\delta\approx 0$ and (b) $\delta\approx -0.2$. At $\delta\approx 0$, the decay of $G_s$ is exponential for 1C, while it is slower for GM. For $\delta\approx -0.2$, both 1C and GM have the first peaks (corresponding to the vibration of particles around their mean positions) well described by the Gaussian, and the decay of the curves is exponential for 1C, while it is not quite exponential for GM.
}
\label{fig: s3}
\end{center}
\end{figure}

\begin{figure}[ht]
\includegraphics*[width=0.5\textwidth]{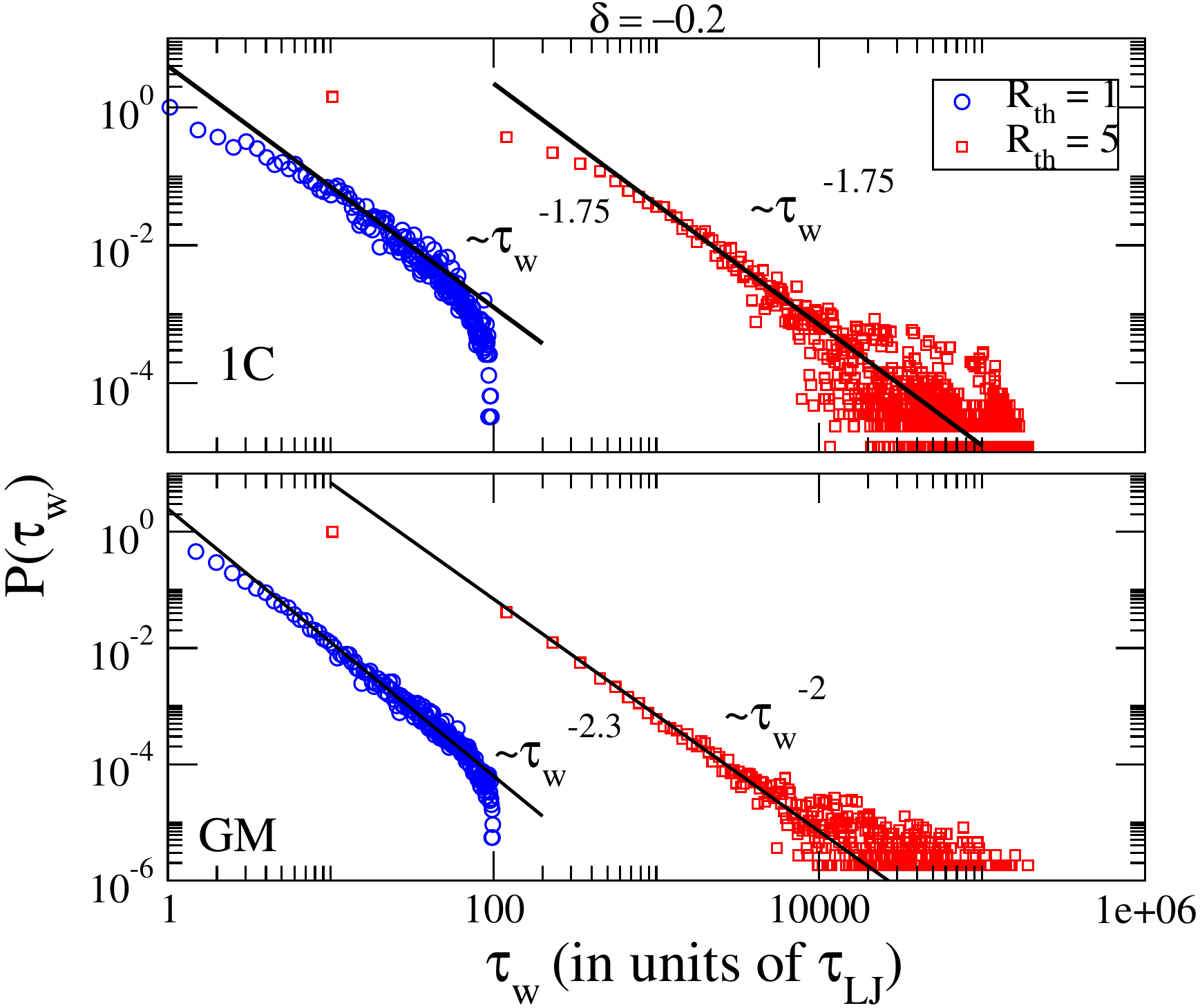}
\begin{center}
\caption{{\bf Waiting time distribution $P(\tau_w)$} at $\delta = -0.2$ (obtained by considering all the particles) is shown for $R_{\rm th}=$ 1 and 5. Having different threshold radius do not affect the exponent $\alpha$ for 1C, while for GM the value of $\alpha$ is slightly lower for larger $R_{\rm th}$.}
\label{fig: s4}
\end{center}
\end{figure}


\begin{thebibliography}{}

\bibitem{SSastry1998_Nature} S. Sastry, P. G. Debenedetti, F. H. Stillinger, {\it Nature}, 393, 554 (1998)
\bibitem{ERWeeks2000_Science} E. R. Weeks {\it et al.}, {\it Science}, 287, 627 (2000)
\bibitem{ERWeeks2002_PRL} E. R. Weeks; and D. A. Weitz, {\it Phys. Rev. Lett.}, 89, 095704 (2002)
\bibitem{Nagel_Nature_2005}  E. I. Corwin, H. M. Jaeger, and S.  R. Nagel, {\it Nature} 435, 1075-1078 (2005).
\bibitem{PinakiC2007_PRL} P. Chaudhuri; L. Berthier; W. Kob, {\it Phys. Rev. Lett.}, 99, 060604 (2007)
\bibitem{HTanaka2010_NatureMaterials} H. Tanaka  {\it et al.}, {\it Nature Materials}, 9, 324 (2010)
\bibitem{JHelfferich2015_PRE1} J. Helffreich {\it et al.}, {\it Phys. Rev. E}, 89, 042603 (2014)
\bibitem{Igal} G. C. Picasso {\it at al.}, {\it J. Chem. Phys.}, 139, 044509 (2013)
\bibitem{AZPatashinski2012_JPCL} A. Z. Patashinski {\it at al.}, {\it J. Phys. Chem. Lett.}, 3, 2431 (2012)

\bibitem{TKawasaki2007_PRL} T. Kawasaki, T. Araki and H. Tanaka, {\it Phys. Rev. Lett.}, 99, 215701 (2007)

\bibitem{Matharoo2006_PRE} G. S. Matharoo, M. S. G. Razul, and P. H. Poole, {\it Phys. Rev. E}, 74, 050502(R) (2006)

\bibitem{CDonati1998_PRL} C. Donati {\it et al.}, {\it Phys. Rev. Lett.}, 80, 2338 (1998)

\bibitem{eli2012_PhysToday} E. Barkai; Y. Garini, and R. Metzler {\it Physics Today}, vol., 29-35 (2012)
\bibitem{MRHorton2010_SoftMatter} M. R. Horton {\it et al.}, {\it Soft Matter}, 6, 2648 (2010)
\bibitem{MJSaxton1997_Biophys} M. J. Saxton, {\it Biophysics Journal}, 72, 1744 (1997)
\bibitem{IIzeddin2014_eLife} I. Izeddin {\it et al.}, {\it eLife}, 3, e02230 (2014)
\bibitem{SRMcGuffee2010_PLoS_ComputBiol} S. R. MsGuffee, and A. H. Elcock, {\it PLoS Comput. Biol.}, 6, e1000694 (2010)
\bibitem{IMSokolov2012_SoftMatter} I. M. Sokolov, {\it Soft Matter}, 8, 9043 (2012)

\bibitem{RMetzler2000_PhysicsReports} R. Metzler, and J. Klafter, {\it Physics Reports}, 339, 1-77 (2000)

\bibitem{allen} M. P. Allen; and D. J. Tildesley, {\it Computer simulation of liquids.} (1987)
\bibitem{lammps} S. J. Plimpton, {\it J. Comput. Phys.} 1995, 117, 1-19.
\bibitem{Shagolsem2015_JCP} L. S. Shagolsem {\it et al.}, {\it J. Chem. Phys.}, 142, 051104 (2015)

\bibitem{mitus} A. C. Mitus {\it et al.}, {\it Phys. Rev. B}, 66, 184202 (2002).

\bibitem{patashinsky}A. Z. Patashinsky {\it et al.}, {\it J. Phys. Chem. C}, 114,  20749 (2010).
\bibitem{barkai} J.-H. Jeon {\it et al.}, {\it Phys. Rev. Lett. } 106, 048103 (2011).
\bibitem{manzo} C. Manzo {\it et al.}, {\it Phys. Rev. X} 5, 011021 (2015)
\bibitem{SBurov2013_PNAS} S. Burov {\it et al.}, {\it PNAS}, 110, 19689 (2013)


\bibitem{Bouchaud1990_PhysRep} J. P. Bouchaud, A. Georges (1990). {\it Physics reports}, 195(4), 127-293.

\bibitem{AVWeigel2011_PNAS} A. V. Weigel {\it et al.}, {\it PNAS}, 108, 6438 (2011)

\bibitem{RMetzler2014_PCCP} R. Metzler {\it et al.}, {\it Phys. Chem. Chem. Phys.}, 16, 24128 (2014)

\bibitem{Benichou} O. Benichou {\it et al.}, {\it Nature Chemistry}, 2, 472 (2010)

\bibitem{dino} D. Osmanovic and Y. Rabin, {\it J. Stat. Phys.}, submitted


\end{thebibliography}
\end{document}